\documentclass[10pt,reqno]{amsart}
\usepackage{amssymb,mathrsfs,xcolor}
\usepackage{pinlabel}

\usepackage{amsmath, amsfonts, amsthm, verbatim}
\usepackage{epstopdf, mathtools}
\mathtoolsset{showonlyrefs}
\usepackage{esint}
\usepackage{stmaryrd}
\usepackage{placeins}

\usepackage{graphicx}
\usepackage{tikz}
\usetikzlibrary{decorations.pathmorphing}

\usepackage{empheq}
\usepackage[shortlabels]{enumitem}

\newcommand{\rar}{\rightarrow}

\DeclareMathOperator{\sech}{sech}
\DeclareMathOperator{\artanh}{artanh}

\usepackage{fullpage}

\newcommand{\ka}{\kappa}
\newcommand{\x}{\xi}

\newcommand{\p}{\partial}

\usepackage[multiple]{footmisc}
\numberwithin{equation}{section}

\newtheorem{thm}{Theorem}[section]

\newtheorem{prop}[thm]{Proposition}

\theoremstyle{remark}

\usepackage{tensor}

\usepackage{slashed}
\usepackage{cite}
\usepackage[hidelinks]{hyperref}

\newif\ifshowrevisions
\showrevisionstrue
\definecolor{revisionblue}{RGB}{0,90,170}
\ifshowrevisions
	
	\newenvironment{revision}{\begingroup\color{revisionblue}}{\endgroup}
\else

\fi
\hypersetup{
	pdftitle={One-dimensional second-gradient elasticity: pantograph-inspired energies and solitary waves},
	pdfauthor={Richard Scott Hallyburton; Ivan Giorgio; Casey Rodriguez}
}

\newcommand{\bbR}{\mathbb R}

\allowdisplaybreaks[3]

\title{One-dimensional second-gradient elasticity: pantograph-inspired energies and solitary waves}

\author{R. S. Hallyburton, I. Giorgio, C. Rodriguez}

\begin{document}

\begin{abstract}
We present a self-contained formulation of longitudinal motion in one-dimensional second-gradient continua. Starting from finite-strain kinematics, we then apply the Principle of Virtual Power to identify the governing field equation and contact actions, derive the balance laws, and establish the mechanical-energy identity for hyperelastic materials. Motivated by pantographic microstructures, we study a class of
finite-strain energies formulated in terms of normalized strain measures
and deformation-dependent strain-gradient moduli. For the associated
equation of motion, a traveling-wave first integral yields a turning-point criterion
for the existence of solitary waves with positive stretch, exponential
decay, and finite mechanical energy, as well as uniqueness at each fixed
speed within a prescribed strain interval. For suitable spring energies, a change of strain variable reduces this integral to a polynomial equation, yielding explicit tensile and compressive profiles for Green--Lagrange and Hencky strains under appropriate parameter and kinematic conditions. Under the positivity condition on the quartic potential, the constructed Hencky energy has several physically natural properties: it is
nonnegative, admits the natural state as its unique zero and global
minimizer, has a positive linearized modulus, penalizes stretch
gradients, and diverges under both material collapse and unlimited
extension. Furthermore, the stored energy is coercive with respect to the $H^1$ norm of logarithmic strain. Together with mechanical-energy conservation, this yields uniform upper and positive lower bounds on stretch throughout the lifespan of smooth solutions.

The theory’s accessible mechanical origin, relatively simple mathematical foundation, and diverse solitary waves offer a tractable entry point to studying the rich nonlinear dispersive equations governing higher-gradient elastic continua.
\end{abstract} 

\maketitle

\section{Introduction}%
\label{sec:introduction}

\subsection{Higher-gradient continua and mathematical motivation}

Generalized continuum mechanics describes materials with intrinsic length scales and organized microstructures. Piola introduced higher-order deformation gradients in the 1840s \cite{dellIsola2014b}, and the Cosserat brothers introduced independent rotational degrees of freedom in the early 1900s \cite{Cosserat1909}. Toupin \cite{Toupin1962a, Toupin1964} and Mindlin \cite{Mindlin1964, Mindlin1965a} developed systematic couple-stress and strain-gradient theories in the 1960s; classical Cosserat rod theories describe structured continua parameterized by one spatial variable \cite{Antman2005}. The Principle of Virtual Power provides a common mechanical foundation for classical and higher-gradient theories by introducing forces and contact actions through duality with virtual velocities and their gradients. Germain systematized this principle in the 1970s, including the representation of microstructural effects \cite{Germain1972, Germain2020, Germain1973}. The balances of forces and moments then follow as consequences \cite{RodriguezDellIsola2026}.

This framework continues to inform the formulation and numerical treatment of structured continua. The work \cite{GrecoCammarataCastelloCuomo2024} proposes an objective finite element formulation for geometrically nonlinear Cosserat rod theory. For higher-gradient models, \cite{HallyburtonRodriguez2026} develops a direct strain-gradient Cosserat rod theory from the principle of virtual work, while \cite{Rodriguez2024, BalitactacEtAl2026, McAvoyShirani2025} derive reduced energy functionals, governing equations, and boundary conditions for rod, plate, and shell models. These developments further connect variational principles with models of generalized continua and numerical methods.

Metamaterials provide a direct motivation for higher-gradient theories through architectured microstructures. In the \emph{synthesis program}, one designs a microstructure whose homogenized limit has a prescribed response \cite{dellIsola2015b}; advances in microscopic fabrication allow these designs and their continuum descriptions to be tested \cite{dellIsola2019b, golaszewski2019metamaterials, Barchiesi2020a}. A basic example is the \emph{pantographic beam}, a periodic truss of bars connected by ideal pivots \cite{alibert2003truss, Barchiesi2019}, with extensions to more complex mechanisms \cite{rizzi26} and nonlinear springs \cite{giorgio2017dynamics}. Heuristic homogenization yields energies depending on derivatives of stretch and curvature, with discrete energies representing finite-difference stencils of higher-order derivatives \cite{Barchiesi2019}. This structure is closely tied to \emph{floppy modes}, non-rigid deformations that store no energy, such as uniform extension of a pantographic beam \cite{alibert2003truss} and constant bending of a Modified Hart's Antiparallelograms Structure (MHAS) \cite{rizzi26} and ZAPAB linkages \cite{MoschiniMurciaDAnnibale2025}. Such families cannot be represented by a non-degenerate first-gradient energy \cite{seppecher2011linear}.

Complementing synthesis, constitutive identification determines continuum parameters from microscopic interactions. The study \cite{LaValleSoize2024Acoustic} uses the acoustic tensor to identify second-gradient models for particle-based materials with pairwise interactions, while \cite{PlacidiMisraLaValleRodriguez2026} refines the identification of strain-gradient elastic parameters in a granular micromechanical setting.

These developments motivate the mathematical study of higher-gradient continua, with one-dimensional second-gradient elasticity offering a simple starting point. For longitudinal displacement $U(X,t)$, the stretch is $F=1+U_X>0$, and the stored energy depends on $F$ and $H=F_X$. With classical kinetic energy, the evolution equation is second order in time and fourth order in space. The theory retains finite-strain kinematics and boundary actions paired with virtual velocity and its spatial derivative. Here we develop a self-contained formulation from the Principle of Virtual Power and construct energies admitting explicit finite-energy solitary waves. The boundary terms and mechanical-energy identity provide a starting point for initial and boundary value problems, while the waves supply concrete solutions for further analysis.

Nonlinearity and dispersion make \textit{solitary waves} a natural first problem. Microstructured dynamics has been studied for decades \cite{Maugin1999, Engelbrecht2005}, with recent developments for pantographic structures \cite{giorgio2017dynamics, Barchiesi2018wave, Barchiesi2026, Eugster2021, Ciallella2022}. A \emph{solitary wave} is a localized traveling wave that preserves its profile during propagation. Such waves have broad physical significance \cite{MantonSutcliffe2004, MunteanuDonescu2004}, including proposed applications to nondestructive testing \cite{KhusnutdinovaSamsonov2008, KhusnutdinovaTranter2015}, and can be a target in synthesis. Numerical studies of pantographic sheets motivated one-dimensional reductions with nonlinear constitutive laws \cite{Eugster2021, Ciallella2022}. For a monolayered pantographic waveguide obtained by geometrically exact heuristic homogenization, including bar micro-rotation inertia, phase-plane analysis identified conditions for tensile solitary waves and numerical computation supplied their profiles \cite{Barchiesi2026}. These studies raise the question of which constitutive choices admit solitary waves and when their profiles can be obtained explicitly.

\subsection{Constitutive models and solitary waves}

The spring--pantograph chain of \cite{giorgio2017dynamics} connects this question with familiar nonlinear dispersive equations. Heuristic homogenization of the chain (Figure~\ref{fig:pantographic-chain} and Section~\ref{sec:microstructure and stored energy}) gives the stored energy
\begin{gather}
	\label{eq:giorgio_energy}
	\Phi(U_X,U_{XX})=\phi_{\text{NL}}(U_X)+\phi_{\text{pan}}(U_{XX}),\\
	\phi_{\text{NL}}(U_X)=\frac{k_1}{2}U_X^2+\frac{k_3}{4}U_X^4,\quad
	\phi_{\text{pan}}(U_{XX})=\frac{k_2}{2}U_{XX}^2,\nonumber
\end{gather}
where $\cdot_X := \partial_X \cdot$, the two terms describe the nonlinear springs and pantographic substructure, and $k_1,k_2,k_3>0$. The Principle of Virtual Power yields
\begin{align}
	\label{eq:giorgio_wave}
	\rho_0 U_{tt}-k_1 U_{XX}\Bigl[1+\frac{3k_3}{k_1}U_X^2\Bigr]+k_2 U_{XXXX}=0,
\end{align}
with reference density $\rho_0$. For the strain\footnote{Here and in the Boussinesq equation displayed below, lowercase $u$ denotes the engineering strain $U_X=F-1$, not a displacement, as is customary in mechanics; the displacement remains uppercase $U$.} $u=U_X$, differentiating
\eqref{eq:giorgio_wave} with respect to $X$ and introducing the
dimensionless independent variables
\begin{gather}
	x=\sqrt{\frac{k_1}{k_2}}\,X,
	\quad
	\tau=\frac{k_1}{\sqrt{\rho_0k_2}}\,t,
\end{gather}
then relabeling $\tau$ as $t$ gives the generalized good Boussinesq equation
\begin{gather}
	u_{tt}-u_{xx}+(u_{xx}+f(u))_{xx}=0,\quad
	f(u)=-\frac{k_3}{k_1}u^3,
\end{gather}
in the convention of \cite{BonaSachs1988}. This defocusing cubic has the opposite sign to the nonlinearities supporting the solitary waves considered there. The traveling-wave first integral for \eqref{eq:giorgio_wave} excludes nontrivial finite-energy solitary waves approaching the natural state. The authors of \cite{giorgio2017dynamics} instead construct numerical quasi-solitary waves by imposing a truncated profile as boundary forcing. Nonlinearity and dispersion alone therefore do not suffice: the constitutive response is decisive.

In this work, we vary two constitutive choices while allowing the corresponding small-strain coefficients to be preserved: the engineering strain and the polynomial spring energy. The original model uses small deformations of the pantographic microstructure \cite{giorgio2017dynamics}, with energy
\begin{gather}
	\label{eq:giorgio_rewritten}
	\Phi(F-1,H)=\Psi(F,H):=\frac{k_1}{2}(F-1)^2+\frac{k_3}{4}(F-1)^4+\frac{k_2}{2}H^2.
\end{gather}
We extend this description to finite strain by changing the strain measure and spring energy, independently and in concert.

Changing the strain measure introduces a deformation-dependent gradient modulus. We use the normalized Seth--Hill family $(\eta_m)_{m\in\bbR}$ \cite{Seth1964, Hill1968}, with $\eta_m(1)=0$ and $\eta_m'(1)=1$ (see \eqref{seth-hill-strains}). More generally, replacing \(F-1\) by a normalized strain measure \(\eta(F)\) and \(H\) by \(\partial_X\eta(F)=\eta'(F)H\) gives
\begin{align}
	\Psi_\eta(F,H)=\mathscr U(\eta(F))+\frac{\kappa_\eta}{2}\bigl(\eta'(F)H\bigr)^2,\quad
	\kappa_\eta>0,
\end{align}
where $\mathscr U$ is the spring energy in the chosen strain. Engineering, Green--Lagrange, and Hencky strains correspond to $m=1,2,0$, respectively. For spring energy quadratic in strain, a turning point argument gives conditions for solitary waves in the Green--Lagrange and Hencky cases. Thus, changing the strain measure can alter solitary-wave existence without altering the lower-order linearized modulus; when $\kappa_\eta$ is chosen equal to the original gradient coefficient, it also preserves the gradient part of the linearized response.

For the governing equation of motion associated with this energy class, the traveling-wave first integral gives a direct criterion for solitary-wave existence: under a subsonic speed condition, its effective potential must be positive between the natural state and a simple turning point within the admissible strain range. Proposition~\ref{general} applies the classical quadrature argument to establish positive stretch, exponential decay, finite mechanical energy, and uniqueness within the prescribed strain segment at a fixed speed. 

To make the quadrature explicitly invertible, we introduce a transformation $r=R(\eta(F))$, determined by $\eta$ alone, and choose a corresponding spring energy so that
$
	\frac{\kappa_\eta}{2}(r')^2=\mathscr V(r)-\frac{\rho_0c^2}{2}r^2,
$
where $c$ is the wave speed and the prime denotes differentiation in the traveling coordinate. This construction isolates the speed dependence in a quadratic term while preserving the stress-free natural state and prescribed linear modulus. Under suitable parameter conditions, a quartic $\mathscr V$ yields an explicit homoclinic profile, whose inverse transformation gives a tensile or compressive solitary wave when the strain remains admissible. For engineering strain, the evolution equation is a generalized good Boussinesq equation with quadratic--cubic nonlinearity. The Green--Lagrange and Hencky equations lie outside this class, yet their profiles remain explicit, with the Green--Lagrange profile expressed through the Lambert $W$ function. The general existence criterion and the range of each strain transformation together identify the conditions for these explicit waves to have positive stretch and finite mechanical energy.

The constructed Hencky model combines explicit waves with useful global energy properties. The geometric characterization of Hencky strain by Neff, Eidel and Martin \cite{neff2016geometry} reduces in one dimension to the invariant distance $|\log F|$ on $\bbR^+$ \cite[Sec.~5]{martin2016minimal}. It is also the only Seth--Hill strain mapping $(0,\infty)$ onto $\bbR$. Under \eqref{positive-transformed-potential-condition}, the Hencky energy constructed from the quartic potential has the following physically useful global energy properties: it is nonnegative, vanishes only at the natural state, has positive linearized modulus, penalizes stretch gradients, and diverges as $F\to0^+$ or $F\to\infty$. The corresponding constructed Green--Lagrange energy, like \eqref{eq:giorgio_rewritten}, remains finite at collapse.  In fact, we demonstrate in Proposition~\ref{prop:stored_energy_coercivity} that, with an admissible quartic potential, the total stored energy associated with $\Psi_L$ controls the $H^1$ norm of $\log F$. By Sobolev embedding, this estimate then proves that $F$ is uniformly bounded from above and away from zero throughout the lifespan of a sufficiently regular, energy-conserving solution.

\subsection{Overview and main results}
The connection between our constitutive energies and discrete microstructures remains open. The energies are pantograph-inspired: we extend the small-strain cell energy without computing the finite kinematics of the bars (cf.\ \cite{Barchiesi2019, Barchiesi2020a, Barchiesi2026}). Designing microstructures that realize these energies and rigorously justifying their continuum limits are corresponding synthesis and homogenization problems. Classical kinetic energy keeps the present model simple; higher-order inertia provides a further extension to be explored in future work \cite{Toupin1964, Germain1973, Barchiesi2026}. The one-dimensional formulation allows these modeling choices and their analytical consequences to be examined directly.

This paper offers an accessible entry point to the mathematical study of higher-gradient continua through a model with relatively simple mechanical foundations and a transparent mathematical structure, whose governing equations are nonlinear dispersive PDEs. The governing equations, contact actions, and mechanical-energy identity provide a framework for investigating well-posedness on the whole line and on bounded intervals, while the explicit solitary waves furnish concrete profiles for stability analysis, perturbation theory, and numerical approximation. Together, the formulation and solutions connect the mechanics of higher-gradient continua with substantial open problems in nonlinear dispersive PDEs.

The paper proceeds from a self-contained mechanical formulation of one-dimensional second-gradient continua to constitutive models and explicit solitary waves. Section 2 develops the finite-strain kinematics and applies the Principle of Virtual Power to identify the governing equation and endpoint contact actions, then establishes the mechanical-energy identity for hyperelastic continua. Section 3 recalls the spring--pantograph chain and constructs continuum energies using normalized finite-strain measures. Section 4 derives the traveling-wave first integral for the associated equation of motion and gives a general turning-point criterion for admissible solitary waves. It first applies the criterion to quadratic spring energies in Green--Lagrange and Hencky strains, then introduces the strain transformation and corresponding spring energies, constructs explicit engineering, Green--Lagrange, and Hencky profiles, and identifies their parameter restrictions. Section 5 discusses open mathematical questions and mechanical extensions. Appendix~\ref{sec:appendixA} proves Proposition~\ref{general}, including uniqueness, finite mechanical energy, and the displacement offset. Appendix~\ref{sec:AppendixB} proves the coercivity estimate and a priori bounds of Proposition~\ref{prop:stored_energy_coercivity}.

\subsection*{Acknowledgements} C. R. gratefully acknowledges support of NSF DMS-2307562. R. S. H. gratefully acknowledges support of the DoD NDSEG Fellowship. R. S. H. also acknowledges that this material is based upon work supported by the Air Force Office of Scientific Research under award number FA9550-25-C-B010 in the amount of \$133,200.

\subsection*{AI Disclosure} ChatGPT 5.6-Sol and Claude Opus 5.5 were used as adversarial readers and for literature review, typo checking, and improvements to the exposition. The authors retain exclusive responsibility for the correctness, accuracy, and validity of the results presented in this work.

\section{One-dimensional second-gradient elasticity}
\label{sec:one-dimensional-second-gradient elasticity}
\subsection{Kinematics}

Let $I \subseteq \mathbb R$ be an interval, the reference configuration of a
one-dimensional continuum, and let $[0,T]$ be a time interval under
consideration. A \textit{motion} is a sufficiently smooth mapping
\begin{align}
	\chi:I\times[0,T]&\to\bbR,\notag\\
	(X,t)&\mapsto x=\chi(X,t),
\end{align}
where $X$ labels a material particle and $x$ is its position at time $t$.
We assume that $\chi(\cdot,t)$ is injective and orientation preserving, so
that
\begin{align}
	F(X,t):=\p_X\chi(X,t)>0.
\end{align}
The \textit{referential displacement} is defined to be
\begin{align}
	U(X,t) := \chi(X,t) - X. 
\end{align}

Throughout this work, a superposed dot denotes the material time
derivative, namely, the time derivative following a fixed material particle.
The material velocity is therefore
\begin{align}
	V(X,t):=\dot\chi(X,t) = \p_t \chi(X,t) = \dot U(X,t).
\end{align}

\begin{figure}[t]
	\centering
	\begin{tikzpicture}[scale=1.0,>=latex]
		\draw[->] (0,2.2) -- (9.2,2.2) node[right] {$X$};
		\draw[very thick] (1.4,2.2) -- (3.9,2.2);
		\fill (1.4,2.2) circle (1.7pt);
		\fill (3.9,2.2) circle (1.7pt);
		\draw[dashed] (1.4,2.2) -- (1.4,2.85);
		\draw[dashed] (3.9,2.2) -- (3.9,2.85);
		\draw[<->] (1.4,2.7) -- node[above] {$\Delta X$} (3.9,2.7);
		\node[below] at (1.4,2.2) {$X$};
		\node[below] at (3.9,2.2) {$X+\Delta X$};
		\node[left] at (0,2.65) {reference};
		
		\draw[->] (0,0) -- (9.2,0) node[right] {$x$};
		\draw[very thick] (4.6,0) -- (8.0,0);
		\fill (4.6,0) circle (1.7pt);
		\fill (8.0,0) circle (1.7pt);
		\draw[dashed] (4.6,0) -- (4.6,0.65);
		\draw[dashed] (8.0,0) -- (8.0,0.65);
		\draw[<->] (4.6,0.5) -- node[above] {$\Delta x$} (8.0,0.5);
		\node[below] at (4.6,0) {$\chi(X,t)$};
		\node[below] at (8.0,0) {$\chi(X+\Delta X,t)$};
		\node[left] at (0,0.45) {current};
		
		\draw[->,thick] (1.4,1.95) to[bend right=18]
		node[above,sloped] {$\chi(\,\cdot\,,t)$} (4.6,0.25);
		\draw[->,thick] (3.9,1.95) to[bend right=12] (8.0,0.25);
	\end{tikzpicture}
	\caption{A material element in the reference and current configurations.}
	\label{fig:one-dimensional-motion}
\end{figure}
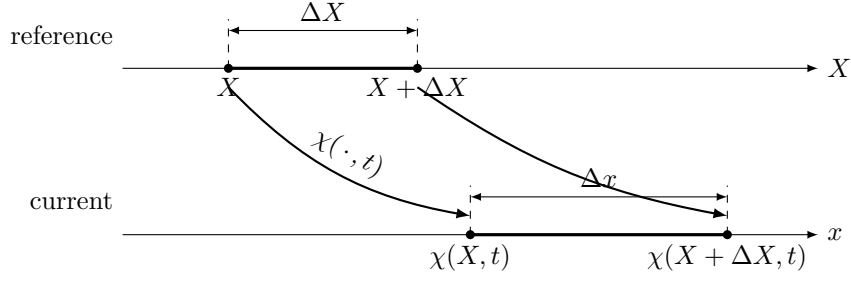

Consider the material interval with endpoints $X$ and $X+\Delta X$. At
time $t$, its endpoints occupy the positions $\chi(X,t)$ and
$\chi(X+\Delta X,t)$, and hence its current length is
\begin{align}
	\Delta x=\chi(X+\Delta X,t)-\chi(X,t).
\end{align}
This correspondence is illustrated in Figure~\ref{fig:one-dimensional-motion}.

The extension of the element relative to its reference length satisfies
\begin{align}
	\lim_{\Delta X\to0}
	\frac{\Delta x-\Delta X}{\Delta X}
	=F(X,t)-1=\p_XU(X,t).
\end{align}
Thus, the displacement gradient \(\p_XU\) is the one-dimensional \textit{engineering strain}, while \(F\) is the deformation gradient, or stretch. The corresponding one-dimensional \textit{Green--Lagrange strain} is obtained by comparing squared lengths:
\begin{align}
	E(X,t)
	&=\frac12\lim_{\Delta X\to0}
	\frac{(\Delta x)^2-(\Delta X)^2}{(\Delta X)^2}
	\notag\\
	&=\frac12\bigl(F(X,t)^2-1\bigr)
	=\p_XU(X,t)
	+\frac12\bigl(\p_XU(X,t)\bigr)^2.
\end{align}
The one-dimensional \textit{Hencky, or logarithmic, strain} is obtained by taking the logarithm of the local length ratio:
\begin{align}
	L(X,t)
	&=\lim_{\Delta X\to0}
	\log\left(\frac{\Delta x}{\Delta X}\right)
	=\log F(X,t).
\end{align}
The condition \(F>0\) ensures that \(L\) is well defined. Unlike the Green--Lagrange strain, whose range is \((-1/2,\infty)\), the Hencky strain maps positive stretches onto the whole real line. In particular, \(L\to-\infty\) as \(F\to0^+\) and \(L\to+\infty\) as \(F\to+\infty\).

The engineering, Green--Lagrange, and Hencky strains belong
to the one-dimensional \textit{Seth--Hill family of strains} \cite{Seth1964,Hill1968}, parametrized by $m\in\mathbb R$,
\begin{align}
	\eta_m(F)
	=
	\begin{cases}
		\dfrac{F^m-1}{m},&m\neq0,\\[1ex]
		\log F,&m=0.
	\end{cases}
	\label{seth-hill-strains}
\end{align}
Indeed,
\begin{align}
	\eta_1(F)=F-1,
	\quad
	\eta_2(F)=\frac12(F^2-1)=E,
	\quad
	\eta_0(F)=\log F=L.
\end{align}
The logarithmic strain arises continuously from this family since
\begin{align}
	\lim_{m\to0}\frac{F^m-1}{m}=\log F.
\end{align}
Every member is normalized by
\begin{align}
	\eta_m(1)=0,
	\quad
	\eta_m'(1)=1,
\end{align}
and therefore has the same linearization about the reference state. More precisely, with \(F=1+\p_XU\),
\begin{align}
	\eta_m(F)
	=\p_XU+\frac{m-1}{2}(\p_XU)^2
	+O\bigl((\p_XU)^3\bigr).
\end{align}
The different values of \(m\) consequently describe the same infinitesimal strain but different finite-strain responses.

For an orientation-preserving one-dimensional continuum, \(F\) and the strains \(\eta_m(F)\) provide equivalent scalar descriptions of the deformation. A superposed proper rigid motion in one dimension consists only of a translation and leaves these quantities unchanged. Moreover,
\begin{align}
	\eta_m'(F)=F^{m-1}>0
	\quad\text{for }F>0,
\end{align}
so every Seth--Hill strain is a one-to-one function of the stretch.

For the second-gradient theories considered below, the gradient of a Seth--Hill strain is
\begin{align}
	\p_X\eta_m(F)=F^{m-1}H,
	\quad
	H:=\p_XF=\p_X^2\chi=\p_X^2U.
	\label{seth-hill-strain-gradient}
\end{align}
In particular,
\begin{align}
	\p_X\eta_1(F)
	&=H,\notag\\
	\p_X\eta_2(F)
	&=\p_XE=FH=:G,\notag\\
	\p_X\eta_0(F)
	&=\p_X\log F=\frac{H}{F}.
\end{align}
Since \(F>0\), each of these strain gradients vanishes precisely when \(H=0\). They therefore provide different finite-strain measures of the spatial variation of stretch, and all vanish for homogeneous deformations.

\subsection{Virtual power and contact actions}

Fix a time $t$ and let $J=(a,b)$ be an arbitrary subinterval of
the reference configuration. A \textit{virtual velocity} is a sufficiently smooth
scalar field $w=w(X)$ on $J$. For a second-gradient continuum, we
assume that the \textit{internal virtual power} depends linearly on the first and
second gradients of $w$ and takes the form
\begin{align}
	\mathscr P_{\rm int}(J;w)
	=
	\int_a^b
	\bigl(P\p_Xw+M\p_X^2w\bigr)\,dX.
	\label{internal-virtual-power}
\end{align}
Here $P=P(X,t)$ is the \textit{Piola force} (the analog of the three-dimensional \textit{Piola stress}), and
$M=M(X,t)$ is the \textit{Piola double force}, dual to the
second gradient of the virtual velocity.

Let $B_{\mathrm{tot}}=B_{\mathrm{tot}}(X,t)$ be a body force per unit reference length. In addition to
the usual endpoint forces $\tau_a$ and $\tau_b$, a second-gradient theory
admits endpoint double forces $\mu_a$ and $\mu_b$, which expend power on
the gradient of the virtual velocity. Accordingly, the \textit{external virtual
power} is defined to be
\begin{align}
	\mathscr P_{\rm ext}(J;w)
	=\int_a^b B_{\mathrm{tot}} w\,dX
	+\tau_aw(a)+\tau_bw(b)
	-\mu_a\p_Xw(a)+\mu_b\p_Xw(b).
	\label{external-virtual-power}
\end{align}
At this stage, $B_{\mathrm{tot}}$ is the total body-force density entering the virtual-power
statement. Its decomposition into applied and inertial parts will be
discussed subsequently.

We postulate the \textit{Principle of Virtual Power} \cite{Germain2020, Germain1973} in the form
\begin{align}
	\mathscr P_{\rm int}(J;w)
	=
	\mathscr P_{\rm ext}(J;w)
	\label{principle-of-virtual-power}
\end{align}
for every subinterval $J \subseteq I$ and every virtual
velocity $w$. Integration by parts twice gives
\begin{align}
	\mathscr P_{\rm int}(J;w)
	=\int_a^b\bigl(-\p_XP+\p_X^2M\bigr)w\,dX
	+\Bigl[(P-\p_XM)w+M\p_Xw\Bigr]_{a}^{b}.
	\label{localized-internal-power}
\end{align}
Since $w$ is arbitrary in the interior of $J$, localization of
\eqref{principle-of-virtual-power} yields the governing field equation
\begin{align}
	\p_X\bigl(P-\p_XM\bigr)+B_{\mathrm{tot}}=0
	\quad\text{in }I.
	\label{second-gradient-equation-of-motion}
\end{align}
The independence of the values of $w$ and $\p_Xw$ at the endpoints gives
the contact actions
\begin{gather}
	\tau_a=-\bigl(P-\p_XM\bigr)(a,t),
	\quad
	\tau_b=\bigl(P-\p_XM\bigr)(b,t),
	\label{endpoint-tractions}\\
	\mu_a=M(a,t),
	\quad \mu_b=M(b,t).
	\label{endpoint-double-forces}
\end{gather}
Equivalently, if $N_{J}$ denotes the outward unit normal to
$J$, so that $N_{J}(a)=-1$ and
$N_{J}(b)=1$, then
\begin{align}
	\tau=\bigl(P-\p_XM\bigr)N_{J},
	\quad
	\mu=M = N_{J} M N_{J}
	\quad\text{on }\p J.
\end{align}
The preceding formulas are the one-dimensional counterparts of those in the full three-dimensional theory. Thus, $P-\p_XM$ is the effective tension that produces the ordinary endpoint
force, while $M$ produces the higher-order endpoint force. No constitutive
restriction on $P$ or $M$ is required thus far; such restrictions
enter when the second-gradient elastic response is specified.

\subsection{Inertia, body forces, and the equation of motion}

We now decompose the total body-force density appearing in
\eqref{second-gradient-equation-of-motion} as
\begin{align}
	B_{\mathrm{tot}}(X,t)=B(X,t)-\rho_0(X)\ddot U(X,t),
	\label{body-force-decomposition}
\end{align}
where $B$ is the \textit{noninertial external body force} per unit reference length, $\rho_0$ is the mass per unit reference length, and
$-\rho_0\ddot U$ is the \textit{inertial body force} per unit reference length. Since
$\ddot\chi=\ddot U$, substitution of
\eqref{body-force-decomposition} into
\eqref{second-gradient-equation-of-motion} gives the dynamic equation
\begin{align}
	\rho_0(X)\ddot U(X,t)
	=
	\p_X\bigl(P(X,t)-\p_XM(X,t)\bigr)+B(X,t).
	\label{dynamic-equation}
\end{align}
Equivalently, in terms of the material velocity $V=\dot U$,
\begin{align}
	\rho_0\dot V=\p_X\bigl(P-\p_XM\bigr)+B.
\end{align}

For the entire body, the natural endpoint conditions associated with
\eqref{dynamic-equation} take the form
\begin{align}
	N_{I}\bigl(P-\p_XM\bigr)=\overline\tau, \quad 
	M=\overline\mu,
\end{align}
on the relevant parts of $\partial I$, where $\overline\tau$ and
$\overline\mu$ are prescribed endpoint forces and double forces. These may
be replaced, independently at each endpoint, by prescribing
conditions on $U$ and $\p_XU$.

\subsection{Balance of linear momentum}

The \textit{linear momentum} of the material subinterval \(J=(a,b)\) is
\begin{align}
	\mathscr L(J,t)
	:=\int_a^b\rho_0(X)V(X,t)\,dX.
\end{align}
To obtain its balance, choose an arbitrary translational virtual velocity (i.e., a generator of uniform translation)
\begin{align}
	w(X)=w_0,
	\quad w_0\in\mathbb R.
\end{align}
Since \(\p_Xw=\p_X^2w=0\), the internal virtual power and the power expended by the endpoint double forces vanish. Using the decomposition
\begin{align}
	B_{\rm tot}=B-\rho_0\dot V
\end{align}
in the principle of virtual power, and recalling that \(w_0\) is arbitrary, gives
\begin{align}
	\frac{d}{dt}\mathscr L(J,t)
	=
	\int_a^bB(X,t)\,dX+\tau_a+\tau_b.
	\label{balance-of-linear-momentum}
\end{align}
Thus, the rate of change of linear momentum equals the resultant applied force. The endpoint double forces do not contribute because they expend no power on a translational virtual velocity.

\subsection{Kinetic-energy identity}

The \textit{kinetic energy} of $J$ is
\begin{align}
	\mathscr K(J,t)
	:=\frac12\int_a^b\rho_0(X)V(X,t)^2\,dX.
\end{align}
Multiplying \eqref{dynamic-equation} by $V$, integrating over $(a,b)$,
and integrating the force terms by parts gives
\begin{align}
	\frac{d}{dt}\mathscr K(J,t)
	&+\int_a^b\bigl(P\p_XV+M\p_X^2V\bigr)\,dX\notag\\
	&=\int_a^bBV\,dX
	+\Bigl[(P-\p_XM)V+M\p_XV\Bigr]_a^b.
	\label{kinetic-energy-theorem}
\end{align}
In view of \eqref{endpoint-tractions} and
\eqref{endpoint-double-forces}, this may be written as
\begin{align}
	\frac{d}{dt}\mathscr K(J,t)
	+\mathscr P_{\rm int}(J;V)
	=\int_a^bBV\,dX
	+\tau_aV(a)+\tau_bV(b)
	-\mu_a\p_XV(a)+\mu_b\p_XV(b).
	\label{kinetic-energy-power-balance}
\end{align}

\subsection{Stored energy and constitutive relations}

A one-dimensional \textit{second-gradient elastic material} is characterized by a
stored-energy density $\Psi=\Psi(F,H;X)$. The internal virtual power is postulated to be the first variation of the total stored energy with respect to the deformation in the direction of the virtual velocity $w$. Thus, for any subinterval $J=(a,b)$,
\begin{align}
	\mathscr P_{\rm int}(J;w)
	:=
	\int_a^b
	\left(
	\p_F\Psi\,\p_X w
	+
	\p_H\Psi\,\p_X^2 w
	\right)\,dX.
	\label{eq:second-gradient-power}
\end{align}
This is equivalent to positing the \textit{constitutive relations} 
\begin{align}
	P=\p_F\Psi,
	\quad
	M=\p_H\Psi.
\end{align}
Combining
\eqref{eq:second-gradient-power} with
\eqref{kinetic-energy-theorem} gives the \textit{balance of total mechanical energy}:
\begin{align}
	\frac{d}{dt}\int_a^b
	\left(\frac12\rho_0V^2+\Psi\right)\,dX
	=\int_a^bBV\,dX
	+\tau_aV(a)+\tau_bV(b)
	-\mu_a\p_XV(a)+\mu_b\p_XV(b).
	\label{global-energy-balance}
\end{align}
In particular, if $B=0$ on $J$ and
$\tau_a=\tau_b=\mu_a=\mu_b=0$, so that no external body force,
traction, or double force acts on the segment $J$, then its total
mechanical energy is conserved:
\begin{align}
	\frac{d}{dt}\int_a^b
	\left(\frac12\rho_0V^2+\Psi\right)\,dX
	=0.
	\label{global-energy-cons}
\end{align}

\subsection{Equation of motion in displacement form}
We henceforth assume that the continuum is homogeneous. Thus, the reference
mass density $\rho_0>0$ is constant and the stored-energy density has no
explicit dependence on the material coordinate:
\begin{align}
	\Psi=\Psi(F,H).
\end{align}
Since
\begin{align}
	F=1+\p_XU,
	\quad
	H=\p_X^2U,
\end{align}
the constitutive relations become
\begin{align}
	P(X,t)&=\Psi_F\bigl(1+\p_XU,\p_X^2U\bigr),\notag\\
	M(X,t)&=\Psi_H\bigl(1+\p_XU,\p_X^2U\bigr).
	\label{constitutive-relations-in-U}
\end{align}
Here and below, subscripts on $\Psi$ denote partial derivatives with respect
to the indicated arguments. Unless an evaluation point is displayed, all
such derivatives are evaluated at
\begin{align}
	(F,H)
	=
	\bigl(1+\p_XU(X,t),\p_X^2U(X,t)\bigr).
	\label{energy-evaluation-point}
\end{align}

Applying the chain rule gives
\begin{align}
	\p_X\Psi_F
	=
	\Psi_{FF}\p_X^2U
	+
	\Psi_{FH}\p_X^3U.
	\label{expanded-derivative-P}
\end{align}
Similarly,
\begin{align}
	\p_X\Psi_H
	=
	\Psi_{FH}\p_X^2U
	+
	\Psi_{HH}\p_X^3U.
\end{align}
Thus, the equation of motion \eqref{dynamic-equation} may be written as 
\begin{align}
    \rho_0 \ddot U = \p_X \left ( \Psi_F - \Psi_{HH} \p_X^3 U - \Psi_{FH} \p_X^2 U \right ) + B. \label{eq:motion}
\end{align}

The equation may be expanded further. A second differentiation yields
\begin{align}
	\p_X^2\Psi_H
	={}&
	\Psi_{FFH}\bigl(\p_X^2U\bigr)^2
	+
	2\Psi_{FHH}\p_X^2U\p_X^3U
	+
	\Psi_{HHH}\bigl(\p_X^3U\bigr)^2\notag\\
	&+
	\Psi_{FH}\p_X^3U
	+
	\Psi_{HH}\p_X^4U.
	\label{expanded-second-derivative-M}
\end{align}
Substitution into \eqref{dynamic-equation} cancels the two terms containing
$\Psi_{FH}\p_X^3U$ and gives the fully expanded equation of motion
\begin{align}
	\rho_0\ddot U
	={}&
	-\Psi_{HH}\p_X^4U
	-\Psi_{HHH}\bigl(\p_X^3U\bigr)^2
	-2\Psi_{FHH}\p_X^2U\p_X^3U\notag\\
	&-\Psi_{FFH}\bigl(\p_X^2U\bigr)^2
	+\Psi_{FF}\p_X^2U
	+B.
	\label{fully-expanded-dynamic-equation}
\end{align}
Together with the initial conditions
\begin{align}
	U(X,0)=U_0(X),
	\quad
	\dot U(X,0)=V_0(X),
\end{align}
\eqref{eq:motion} or \eqref{fully-expanded-dynamic-equation} defines the formal initial-value problem for a
one-dimensional second-gradient elastic continuum, subject to specified boundary
conditions when $I\neq\bbR$. The fourth-order spatial term in \eqref{fully-expanded-dynamic-equation} is the characteristic \textit{dispersive}
contribution of the second-gradient theory.

The next section introduces pantograph-inspired discrete interactions and
uses their formal continuum limits to motivate constitutive choices for
the stored energy $\Psi(F,H)$.

\section{One-dimensional pantographic microstructures and stored energies}
\label{sec:microstructure and stored energy}
We now develop a class of one-dimensional second-gradient energies motivated by pantographic microstructures. We begin with a discrete chain of identical masses embedded in a pantographic substructure, with neighboring masses connected by nonlinear springs (see Figure~\ref{fig:pantographic-chain}). We first recall the small-strain quadratic energy assigned to the pantographic substructure in \cite{giorgio2017dynamics}, and then propose a finite-strain constitutive generalization. The same finite-strain measures may also be incorporated into the lower-order spring energy.

The limits below are intended as discrete-to-continuum heuristics. In particular, we do not assert that the generalized energies are the unique homogenized energies associated with a specified bar-and-hinge microstructure. Rather, they retain the basic mechanism of the small-strain discrete model while allowing its finite-strain response to be specified constitutively. The design (or \textit{synthesis} problem) of bar-and-hinge microstructures whose continuum limits yield the proposed finite-strain second-gradient energies remains an open problem for future research. 

\subsection{Discrete kinematics and the pantographic chain}

Let the identical reference masses be $P_i$, with positions $X_i=i\epsilon$, and
write
\[
\chi_i(t)=X_i+u_i(t),
\]
where $\epsilon>0$ is the cell size. The discrete relative displacement and 
discrete stretch are given by
\begin{align}
	\Delta_i u
	:={}&u_{i+1}-u_i,\notag\\
	F_i^\epsilon
	:={}&\frac{\chi_{i+1}-\chi_i}{\epsilon}
	=1+\frac{\Delta_i u}{\epsilon}.
	\label{pantographic-discrete-bond-strains}
\end{align}
We assume that there exists a fixed smooth deformation $\chi$ satisfying
$\chi(X_i,t)=\chi_i(t)$. Taylor expansion yields
	\begin{align}
		F_i^\epsilon
		=F(X_i,t)+O(\epsilon)
		\quad\text{as }\epsilon\to0.
	\end{align}
For later comparison,
define the discrete gradients
\begin{align}
	H_i^\epsilon
	:={}&\frac{F_i^\epsilon-F_{i-1}^\epsilon}{\epsilon}
	=\frac{u_{i+1}-2u_i+u_{i-1}}{\epsilon^2}.
	\label{pantographic-discrete-gradient-strains}
\end{align}
Taylor expansion about $X_i$ similarly gives
	\begin{align}
		H_i^\epsilon=H(X_i,t)+O(\epsilon^2)
		\quad\text{as }\epsilon\to0.
	\end{align}

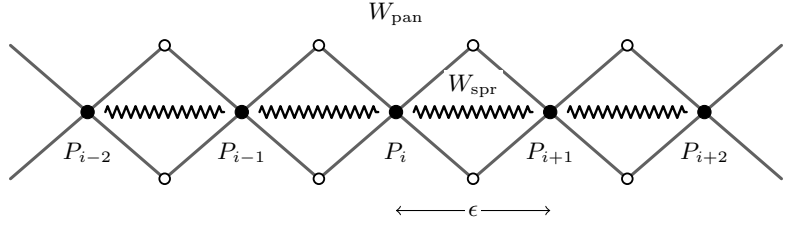
\begin{figure}[t]
	\centering
	\begin{tikzpicture}[
		x=1.02cm,y=0.84cm,
		line cap=round,line join=round,
		every node/.style={text=black,font=\small}]
		\draw[draw=black!62,line width=1.15pt]
		(-1,1.05)--(1,-1.05)--(3,1.05)--(5,-1.05)--
		(7,1.05)--(9,-1.05);
		\draw[draw=black!62,line width=1.15pt]
		(-1,-1.05)--(1,1.05)--(3,-1.05)--(5,1.05)--
		(7,-1.05)--(9,1.05);
		\foreach \x in {1,3,5,7}{
			\filldraw[fill=white,draw=black,line width=0.7pt]
			(\x,1.05) circle (2.0pt);
			\filldraw[fill=white,draw=black,line width=0.7pt]
			(\x,-1.05) circle (2.0pt);
		}
		\foreach \x in {0,2,4,6}{
			\draw[draw=black,line width=0.9pt,decorate,
			decoration={zigzag,segment length=3.5pt,amplitude=2.2pt}]
			({\x+0.23},0)--({\x+1.77},0);
		}
		\foreach \x/\lab in
		{0/{i-2},2/{i-1},4/i,6/{i+1},8/{i+2}}{
			\filldraw[fill=black,draw=black] (\x,0) circle (2.5pt);
			\node[below=8pt] at (\x,0) {$P_{\lab}$};
		}
		\node[above=4pt,fill=white,inner sep=1.5pt] at (5,0)
		{$W_{\rm spr}$};
		\node[above=1pt] at (4,1.18) {$W_{\rm pan}$};
		\draw[draw=black,<->] (4,-1.55)--(6,-1.55)
		node[midway,fill=white,inner sep=1pt] {$\epsilon$};
	\end{tikzpicture}
	\caption{An idealized one-dimensional pantographic structure as in
		\cite{giorgio2017dynamics}. Adjacent masses are connected by
		nonlinear springs, while the two crossing families of inextensible
		bars transmit a pantographic interaction through ideal hinges.}
	\label{fig:pantographic-chain}
\end{figure}

Figure~\ref{fig:pantographic-chain} shows the idealized microstructure
presented in \cite{giorgio2017dynamics}. The masses along the centerline
carry the longitudinal degrees of freedom. Nonlinear springs connect
adjacent masses, while two families of hinged, inextensible flexible
bars form the pantographic substructure.

\subsection{Pantographic contributions}

In the discrete model, the pantographic substructure stores energy when
the extension is nonuniform. For small deflections of the bars, the
stored energy contributed by the $i$th pantographic cell is
\cite{alibert2003truss, seppecher2011linear}
\begin{align}
	\Phi_{{\rm pan},i}^\epsilon
	=\frac{K_\epsilon}{2}
	\bigl(u_{i+1}-2u_i+u_{i-1}\bigr)^2.
	\label{pantographic-discrete-H-energy}
\end{align}
	With the scaling
	\begin{align}
		K_\epsilon=\frac{\ka_H}{\epsilon^3},
		\quad {\ka_H>0},
	\end{align}
	equation \eqref{pantographic-discrete-H-energy} becomes
	\begin{align}
		\Phi_{{\rm pan},i}^\epsilon
		=\epsilon\,\frac{\ka_H}{2}(H_i^\epsilon)^2.
	\end{align}
	Since $H_i^\epsilon$ approximates $H(X_i,t)$, the sum is a Riemann sum, and heuristically, as $\epsilon\to0$,
	\begin{align*}
		\sum_i\Phi_{{\rm pan},i}^\epsilon
		\to \int_I\frac{\ka_H}{2}H^2\,dX.
	\end{align*}

\subsection{Finite-strain generalization of the pantographic interaction}

Equation \eqref{pantographic-discrete-H-energy} can also be written as
\begin{align}
	\Phi_{{\rm pan},i}^\epsilon
	=\epsilon\frac{\ka_H}{2}
	\left(
	\frac{F_i^\epsilon-F_{i-1}^\epsilon}{\epsilon}
	\right)^2.
	\label{pantographic-discrete-H-energy-rewritten}
\end{align}
Thus, the original interaction penalizes differences between the stretches
of neighboring cells. At finite strain, however, stretch is not the only
scalar quantity by which deformation may be measured. Indeed, let
$J\subseteq(0,\infty)$ be an open interval containing $1$, and let
$\eta:J\to\eta(J)\subseteq\mathbb R$ be a smooth strain measure satisfying
$\eta'(F)>0$ for every $F\in J$ and
\begin{align}
	\eta(1)=0,
	\quad
	\eta'(1)=1.
	\label{normalized-finite-strain-measure}
\end{align}
The normalization gives
\begin{align}
	\eta(F)=F-1+O\bigl((F-1)^2\bigr)
	\quad\text{as }F\to1.
	\label{finite-strain-linearization}
\end{align}

Associate the discrete strain
$\eta_i^\epsilon=\eta(F_i^\epsilon)$ with the $i$th position and replace the difference of
neighboring stretches in
\eqref{pantographic-discrete-H-energy-rewritten} by the difference of
their finite strains. The generalized cell energy is
\begin{align}
	\Phi_{{\rm pan},i}^{\epsilon,\eta}
	=\epsilon\frac{\ka_\eta}{2}
	\left(
	\frac{\eta(F_i^\epsilon)-\eta(F_{i-1}^\epsilon)}{\epsilon}
	\right)^2.
	\label{generalized-discrete-pantographic-energy}
\end{align}
{Here $\ka_\eta>0$.}
This interaction couples neighboring cells, is nonnegative, and vanishes
precisely when the strain is constant from one cell to the next. Since
$\eta$ is one-to-one, this is equivalent to the stretch being constant.
It is unchanged by a superposed translation because it depends only on
the discrete stretches. Moreover, \eqref{finite-strain-linearization} shows
that \eqref{generalized-discrete-pantographic-energy} and the original
quadratic interaction have the same leading-order small-strain response
when $\kappa_\eta=\kappa_H$. We therefore regard it as a finite-strain extension that preserves the linearized theory.

For a smooth deformation,
\begin{align}
	\frac{\eta(F_i^\epsilon)-\eta(F_{i-1}^\epsilon)}{\epsilon}
	=[\eta'(F)H](X_i,t) + O(\epsilon).
\end{align}
Consequently,
\begin{align}
	\sum_i\Phi_{{\rm pan},i}^{\epsilon,\eta}
	\to
	\int_I\frac{\ka_\eta}{2}
	\bigl(\eta'(F)H\bigr)^2\,dX.
	\label{generalized-pantographic-continuum-limit}
\end{align}
The corresponding gradient density has the form
\begin{align}
	\Psi_{{\rm grad},\eta}(F,H)
	=\frac{A_\eta(F)}{2}H^2,
	\quad
	A_\eta(F)=\ka_\eta\bigl(\eta'(F)\bigr)^2>0.
	\label{generalized-pantographic-gradient-density}
\end{align}

For the Seth--Hill strain $\eta_m$ introduced in
\eqref{seth-hill-strains}, equation
\eqref{seth-hill-strain-gradient} gives the following expression, where {$\ka_m:=\ka_{\eta_m}>0$}:
\begin{align}
	A_m(F)=\ka_mF^{2m-2}.
\end{align}
In particular, $m=1$, $m=2$, and $m=0$ give the engineering,
Green--Lagrange, and Hencky strain-gradient contributions
\begin{align}
	\frac{\ka_H}{2}H^2,
	\quad
	\frac{\ka_G}{2}G^2
	=\frac{\ka_G}{2}F^2H^2,
	\quad
	\frac{\ka_L}{2}\bigl((\log F)_X\bigr)^2
	=\frac{\ka_L}{2F^2}H^2.
	\label{engineering-green-hencky-gradient-energies}
\end{align}
When their coefficients are equal, these densities have the same
quadratic small-strain approximation, but they differ at finite strain.

Examples of finite-strain energies derived from specified pantographic
substructures are given in \cite{Barchiesi2019,Barchiesi2020a}.

\subsection{Nonlinear-spring contributions}

The second discrete contribution is the energy stored in the nonlinear springs. The model in \cite{giorgio2017dynamics} assigns to the \(i\)th spring the symmetric quadratic-plus-quartic potential
\begin{align}
	\Phi_{{\rm spr},i}^\epsilon
	=\frac{\mu_\epsilon}{2}(\Delta_i u)^2
	+\frac{b_\epsilon}{4}(\Delta_i u)^4,
	\label{pantographic-discrete-spring-energy}
\end{align}
where
\begin{align}
	\mu_\epsilon=\frac{\mu}{\epsilon},
	\quad
	b_\epsilon=\frac{b}{\epsilon^3},
	\quad {\mu>0,\ b>0}.
\end{align}
In terms of $F_i^\epsilon$ we have 
\begin{align}
	\Phi_{{\rm spr},i}^\epsilon
	=\epsilon\left[
	\frac{\mu}{2}(F_i^\epsilon-1)^2
	+\frac{b}{4}(F_i^\epsilon-1)^4
	\right].
\end{align}
Thus, the sum of the discrete spring energies is a Riemann sum. As
\(\epsilon\to0\), we have the heuristic convergence
\begin{align}
	\sum_i\Phi_{{\rm spr},i}^\epsilon
	\to
	\int_I W(F)\,dX,
	\label{pantographic-spring-continuum-limit}
\end{align}
where
\begin{align}
	W(F)
	=\frac{\mu}{2}(F-1)^2
	+\frac{b}{4}(F-1)^4.
\end{align}

This construction extends directly to a general spring energy. Let
\(W:J\to[0,\infty)\) be a fixed function defined on an interval
\(J\subseteq(0,\infty)\), and assume that \(F_i^\epsilon\in J\). Suppose that the $i$-th spring potential takes the form
\begin{align}
	\Phi_{{\rm spr},i}^\epsilon
	=\epsilon W(F_i^\epsilon).
\end{align}
Suppose that $I$ is the union of the cells occurring in the sum. As before, we have
the heuristic convergence
\begin{align}
	\sum_i\Phi_{{\rm spr},i}^\epsilon
	\to
	\int_I W(F)\,dX,
	\quad\text{as }\epsilon\to0.
	\label{pantographic-spring-continuum-limit-general}
\end{align}

The finite-strain measure used in the pantographic interaction may also
be used to describe the deformation of each nonlinear spring. Let
$\mathscr U:\eta(J)\to[0,\infty)$ and define
\begin{align}
	\Phi_{{\rm spr},i}^{\epsilon,\eta}
	=\epsilon\mathscr U\bigl(\eta(F_i^\epsilon)\bigr).
	\label{finite-strain-discrete-spring-energy}
\end{align}
The corresponding continuum limit is
\begin{align}
	\sum_i\Phi_{{\rm spr},i}^{\epsilon,\eta}
	\to
	\int_I\mathscr U\bigl(\eta(F)\bigr)\,dX.
	\label{finite-strain-spring-continuum-limit}
\end{align}
Natural conditions at the reference state are
\begin{align}
	\mathscr U(0)=0,
	\quad
	\mathscr U'(0)=0,
	\quad
	\mathscr U''(0)=\mu>0.
	\label{finite-strain-spring-reference-conditions}
\end{align}
The first condition normalizes the reference energy to zero, the second makes the homogeneous reference state tension-free, and the third prescribes its homogeneous linearized spring modulus. Indeed, using
\eqref{normalized-finite-strain-measure},
\begin{align}
	\frac{d}{dF}\mathscr U(\eta(F))
	\Bigr |_{F=1}
	={}&0,\\
	\frac{d^2}{dF^2}\mathscr U(\eta(F))
	\Bigr |_{F=1}
	={}&\mu.
\end{align}
If $\mathscr U(s)>0$ for $s\neq0$, then $F=1$ is the unique global
minimizer on $J$.

Since $\eta$ is one-to-one, any stretch energy $W(F)$ can formally be
written as
\begin{align}
	W(F)=\mathscr U(\eta(F)),
	\quad
	\mathscr U(s)=W\bigl(\eta^{-1}(s)\bigr).
\end{align}
Thus, allowing a general function of a finite-strain measure does not
restrict the class of lower-order energies. The constitutive distinction
appears when a particular functional form, such as a polynomial, is
prescribed in one strain measure rather than another. A polynomial in
$F-1$, a polynomial in $E$, and a polynomial in $\log F$ generally define
different finite-strain materials even when they have the same
linearization at $F=1$.

\subsection{Combined pantograph-inspired energies}

Using the same strain measure in the gradient and lower-order
contributions gives the matched family
\begin{align}
	\Psi_\eta(F,H)
	=\frac{\ka_\eta}{2}
	\bigl(\eta'(F)H\bigr)^2
	+\mathscr U\bigl(\eta(F)\bigr).
	\label{matched-pantographic-energy-family}
\end{align}
The engineering, Green--Lagrange, and Hencky choices yield\footnote{A subscript $G$ or $L$ does not indicate a partial derivative with respect to that variable.}
\begin{align}
	\Psi_{\rm eng}(F,H)
	={}&\frac{\ka_H}{2}H^2+\mathscr U(F-1),\notag\\
	\Psi_G(F,H)
	={}&\frac{\ka_G}{2}G^2+\mathscr U(E),\notag\\
	\Psi_L(F,H)
	={}&\frac{\ka_L}{2F^2}H^2+\mathscr U(\log F).
	\label{matched-pantographic-examples}
\end{align}

In these three expressions, $\mathscr U$ denotes a stored-energy function
defined on the range of the corresponding strain measure; the same
functional form need not be used in all three models.

There is no general requirement that the same strain measure be used
in both contributions. The nonlinear springs and the pantographic
substructure represent distinct mechanisms, and one may more generally
consider
\begin{align}
	\Psi_{\eta,\xi}(F,H)
	=\frac{\ka_\eta}{2}
	\bigl(\eta'(F)H\bigr)^2
	+\mathscr U\bigl(\xi(F)\bigr),
	\label{mixed-pantographic-energy-family}
\end{align}
where $\eta$ and $\xi$ are independently chosen normalized strain
measures. For example, $H^2+\mathscr U(E)$ gives a mixed energy, while $G^2+\mathscr U(E)$ belongs to the matched family studied below. The matched
family \eqref{matched-pantographic-energy-family} is nevertheless a
particularly natural starting point when both discrete mechanisms are
assumed to measure deformation in the same finite-strain variable.

The affine choice $\eta(F)=F-1$ produces a constant gradient modulus.
Non-affine choices produce the deformation-dependent modulus $A_\eta(F)$
in \eqref{generalized-pantographic-gradient-density}. In the next section,
we obtain explicit finite-energy solitary waves for physically reasonable 
representative energies.

\section{Existence and explicit construction of finite-strain solitary waves}
\label{sec:explicit families}
We develop an existence criterion and an explicit construction for solitary waves associated with the family of energies \eqref{matched-pantographic-energy-family}. We first derive the traveling-wave first integral and identify conditions for an admissible homoclinic orbit. We apply the criterion to quadratic spring energies, then choose spring energies for which a strain transformation makes the profiles explicit, and apply the construction to engineering, Green--Lagrange, and Hencky strains. Throughout this section, let $I=\bbR$, suppose that $B=0$, and consider a traveling deformation
\begin{align}
	z=X-ct,
	\quad
	U(X,t)=\mathcal U(z),
	\quad
	F(X,t)=f(z)=1+\mathcal U'(z),
	\label{finite-strain-traveling-wave-ansatz}
\end{align}
where a prime denotes differentiation with respect to $z$. We seek traveling-wave solutions whose stretch increment is localized.
More precisely, we call such a solution a solitary wave if
\begin{align}
	f(z)\to1, \quad 
	f'(z)\to0, \quad 
	f''(z)\to0 
	\quad\text{as }z\to\pm\infty.
	\label{finite-strain-solitary-wave-conditions}
\end{align}
Thus, the stretch approaches its natural value $F=1$ at both
ends. The displacement need not approach the same constant at the two
ends; rather,
\begin{align}
	U(+\infty,t)-U(-\infty,t)
	=\int_{\bbR}\bigl(f(z)-1\bigr)\,dz,
	\label{finite-strain-displacement-shift}
\end{align}
provided that the integral converges.

\subsection{First integral and an existence criterion}

Let $\eta$ be a smooth strain measure defined on an open interval $J\subseteq(0,\infty)$ containing $1$, satisfying \eqref{normalized-finite-strain-measure} and $\eta'(F)>0$ on $J$. Let $K = \eta(J)$ and
\begin{align}
	\Psi_\eta(F,H)
	=\frac{\ka_\eta}{2}\bigl(\eta'(F)H\bigr)^2
	+\mathscr U\bigl(\eta(F)\bigr),
	\quad \ka_\eta>0,
	\label{solvable-matched-energy}
\end{align}
where $\mathscr U\in C^\infty(K)$ satisfies
\begin{align}
	\mathscr U(0)=0,
	\quad
	\mathscr U'(0)=0,
	\quad
	\mathscr U''(0)=\mu>0.
	\label{solvable-energy-natural-state}
\end{align}
Since $\p_X\eta(F)=\eta'(F)H$, the constitutive relations associated with
\eqref{solvable-matched-energy} are
\begin{align}
	P
	&=\eta'(F)\mathscr U'\bigl(\eta(F)\bigr)
	+\ka_\eta\eta'(F)\eta''(F)H^2,
	\notag\\
	M
	&=\ka_\eta\bigl(\eta'(F)\bigr)^2H.
\end{align}
Differentiating $M$ and collecting terms gives the effective tension in the
form
\begin{align}
	P-\p_XM
	=\eta'(F)
	\left[
	\mathscr U'\bigl(\eta(F)\bigr)
	-\ka_\eta\p_X^2\eta(F)
	\right].
	\label{matched-energy-effective-tension}
\end{align}

For the solitary wave in
\eqref{finite-strain-traveling-wave-ansatz}, set
\begin{align}
	\varepsilon(z)=\eta(f(z)).
\end{align}
Substitution of \eqref{matched-energy-effective-tension} into
\eqref{dynamic-equation}, followed by one integration, gives
\begin{align}
	\eta'(f)
	\left(\mathscr U'(\varepsilon)-\ka_\eta\varepsilon''\right)
	=\rho_0c^2(f-1).
	\label{once-integrated-matched-wave-equation}
\end{align}
The integration constant is fixed by the solitary-wave limits \eqref{finite-strain-solitary-wave-conditions} and the natural-state conditions \eqref{solvable-energy-natural-state}.
Dividing by $\eta'(f)$ and
multiplying by $\varepsilon'$ yields an exact derivative because
\begin{align}
	\frac{f-1}{\eta'(f)}\varepsilon'
	=(f-1)f'
	=\frac{d}{dz}\frac{(f-1)^2}{2}.
\end{align}
A second integration, using the same limits to set the integration constant to zero, gives the first integral
\begin{align}
	\frac{\ka_\eta}{2}(\varepsilon')^2
	=\mathscr U(\varepsilon)
	-\frac{\rho_0c^2}{2}
	\left(\eta^{-1}(\varepsilon)-1\right)^2 := Q_c(\varepsilon). 
	\label{finite-strain-universal-first-integral}
\end{align}

The following proposition applies the classical quadrature construction for homoclinic orbits of scalar conservative equations to the present finite-strain model. For completeness, we include a proof in the appendix to establish the kinematic admissibility, precise decay, and finite mechanical energy of the resulting waves.
\begin{prop}
	\label{general}
	Suppose $\mu_c:=\mu-\rho_0c^2>0$. Suppose there are
	$\sigma\in\{-1,1\}$ and $A>0$ such that the closed segment joining
	$0$ and $\sigma A$ is contained in $K = \eta(J)$ and, with
	$q(s)=Q_c(\sigma s)$,
	\begin{equation}
		\forall s \in (0, A), \, q(s) > 0, \quad q(A)=0,\quad q'(A)<0.
		\label{turning}
	\end{equation}
	Then \eqref{dynamic-equation} has a nonconstant smooth traveling solitary wave with
	positive stretch profile $f > 0$. After translating the profile center to $z=0$, its strain
	$\varepsilon(z)$ is even, has sign $\sigma$, satisfies
	$\varepsilon(0)=\sigma A$, and $|\varepsilon(z)|$ decreases strictly for
	$z>0$. It satisfies \eqref{finite-strain-universal-first-integral} and
	\[
	f(z)\to1,\quad f'(z)\to0,\quad f''(z)\to0
	\quad (z\to\pm\infty).
	\]
	For $\lambda_{\eta,c}=\sqrt{\mu_c/\kappa_\eta}$ there is a constant $C>0$ such that
	\begin{align}
		\varepsilon(z)&=\sigma C e^{-\lambda_{\eta,c}|z|}
		+O(e^{-2\lambda_{\eta,c}|z|}),\label{tails}\\
		f(z)-1&=\sigma C e^{-\lambda_{\eta,c}|z|}+O(e^{-2\lambda_{\eta,c}|z|}).\notag
	\end{align}
	The total mechanical energy is absolutely integrable:
	\begin{equation}
		\int_\mathbb R\left(\frac{\rho_0c^2}{2}(f-1)^2
		+\frac{\kappa_\eta}{2}(\varepsilon')^2+\mathscr{U}(\varepsilon)\right)\,dz
		\quad\hbox{is finite}.\label{total}
	\end{equation}
	The displacement has finite limits at both ends, with nonzero offset of
	sign $\sigma$. Up to translation of $z$ and addition of a constant to
	$\mathcal U$, this is the unique nonconstant classical solitary wave with speed $c$ satisfying $\mu - \rho_0 c^2 > 0$
	whose strain takes values in the segment joining $0$ and $\sigma A$.
\end{prop}

\subsection{Solitary waves for quadratic spring energies}

A first application of Proposition~\ref{general} shows that changing the strain measure alone can produce solitary waves with quadratic spring energy. We retain the quadratic dependence familiar from small-strain elasticity, but express it in a normalized finite-strain measure:
\begin{align}
	\mathscr U(\varepsilon)=\frac{\mu}{2}\varepsilon^2,
	\quad
	\Psi_\eta(F,H)=\frac{\kappa_\eta}{2}
	\bigl(\eta'(F)H\bigr)^2+\frac{\mu}{2}\eta(F)^2.
\end{align}
These energies share the linearized spring modulus $\mu$, while their finite-strain responses depend on $\eta$. To apply the proposition, introduce the normalized speed
\begin{align}
	\theta:=|c|\sqrt{\frac{\rho_0}{\mu}},
\end{align}
so that the subsonic condition is $\theta<1$. We then seek a simple turning point of $Q_c$ within the admissible strain range. This gives compressive waves for Green--Lagrange strain and tensile waves for Hencky strain.
\subsubsection{Green--Lagrange strain}
For the Green--Lagrange strain $\varepsilon=E=(F^2-1)/2$, the stretch increment $s=F-1>-1$ gives $E=s+s^2/2$ and
\begin{align}
	Q_c(E)
	=\frac{\mu}{2}s^2
	\left[\left(1+\frac{s}{2}\right)^2-\theta^2\right].
\end{align}
A nonzero turning point must therefore satisfy
\begin{align}
	s_*=-2(1-\theta),\quad
	F_*=2\theta-1,\quad
	E_*=-2\theta(1-\theta).
\end{align}
Positive stretch requires $\theta>1/2$, while the subsonic condition requires $\theta<1$. For these speeds, set $\sigma=-1$ and $A=2\theta(1-\theta)$. Then $0<A<1/2$, so $[-A,0]$ lies in the admissible strain range $(-1/2,\infty)$. Moreover, $Q_c(E)>0$ for $-A<E<0$, and
\begin{align}
	Q_c'(-A)=\frac{2\mu\theta(1-\theta)^2}{2\theta-1}>0.
\end{align}
Thus $q(s):=Q_c(-s)$ satisfies the turning-point conditions $q(A)=0$ and $q'(A)<0$. Proposition~\ref{general} gives a compressive solitary wave whenever
\begin{align}
	\frac12\sqrt{\frac{\mu}{\rho_0}}
	<|c|<
	\sqrt{\frac{\mu}{\rho_0}}.
\end{align}
Its minimum stretch is $F_*=2\theta-1$, and the stretch increases strictly toward $1$ away from the center. For $0\leq\theta\leq1/2$, there is no nonzero admissible compressive turning point; for $\theta<1$, there is no tensile turning point. The interval above therefore exhausts the nonconstant subsonic solitary waves of the quadratic Green--Lagrange model.

For $1/2 < \theta< 1$, we can describe the solitary wave in terms of an explicit parametric relation. Let $\omega := \sqrt{\frac{1-\theta}{1+\theta}}$, and, for $\xi \in \bbR$, consider the functions
\begin{gather}
\label{eq:GL_parameterization}
    z(\xi) := \sqrt{\frac{\kappa_\eta}{\mu}} \left[ \frac{2\xi}{\sqrt{1- \theta^2 }} - 4 \artanh(\omega \tanh \xi)\right], \quad
    f(\xi) -1 := -\frac{2(1-\theta^2)}{1 + \theta \cosh (2\xi)}.
\end{gather}
A direct computation gives
\[
z'(\xi) = \frac{2}{\sqrt{1-\theta^2}}\sqrt{\frac{\kappa_\eta}{\mu}}\, f(\xi).
\]
Since $z$ is an increasing bijection of $\mathbb{R}$, the strain profile
$\varepsilon(z) := \eta_2\bigl(f(\xi(z))\bigr)$, where $\xi(z)$ denotes the inverse of
$z(\xi)$, is well defined.
A direct computation then shows that $\varepsilon$ satisfies \eqref{finite-strain-universal-first-integral}, with
$f(0)=F_*$, $f$ even, and $f(\xi)\to 1$ as $\xi\to\pm\infty$. By the uniqueness in
Proposition~\ref{general}, the curve $\xi\mapsto\bigl(z(\xi),\,f(\xi)-1\bigr)$ is the
compressive solitary-wave profile.

\subsubsection{Hencky strain}
The same quadratic spring energy gives a different turning-point structure for Hencky strain. With $\varepsilon=L=\log F$,
\begin{align}
	Q_c(L)=\frac{\mu}{2}
	\left[L^2-\theta^2(e^L-1)^2\right].
\end{align}
On the tensile side, define
\begin{align}
	g(L):=\frac{e^L-1}{L},\quad L>0.
\end{align}
Since $g'(L)>0$ and $g$ has range $(1,\infty)$, each $0<\theta<1$ determines a unique $A>0$ through
\begin{align}
	g(A)=\frac1\theta,
	\quad\text{equivalently}\quad
	A=\theta(e^A-1).
\end{align}
The factorization
\begin{align}
	Q_c(L)=\frac{\mu}{2}L^2
	\left[1-\theta^2g(L)^2\right]
\end{align}
then gives $Q_c(L)>0$ for $0<L<A$ and
\begin{align}
	Q_c'(A)=-\mu\theta A^2g'(A)<0.
\end{align}
Proposition~\ref{general}, with $\sigma=1$, therefore yields a tensile solitary wave for every speed satisfying
\begin{align}
	0<|c|<\sqrt{\frac{\mu}{\rho_0}}.
\end{align}
Its maximum stretch is $e^A$, and the stretch decreases strictly toward $1$ away from the center. On the compressive side, $0<(e^L-1)/L<1$, so $Q_c(L)>0$ for all $L<0$ when $\theta<1$ and no turning point occurs. At $c=0$, $Q_c$ is purely quadratic and has no nonzero root. Thus every nonconstant subsonic solitary wave of the quadratic Hencky model is tensile and has nonzero speed.
\begin{figure}[t]
	\centering
	\includegraphics[width=\linewidth]{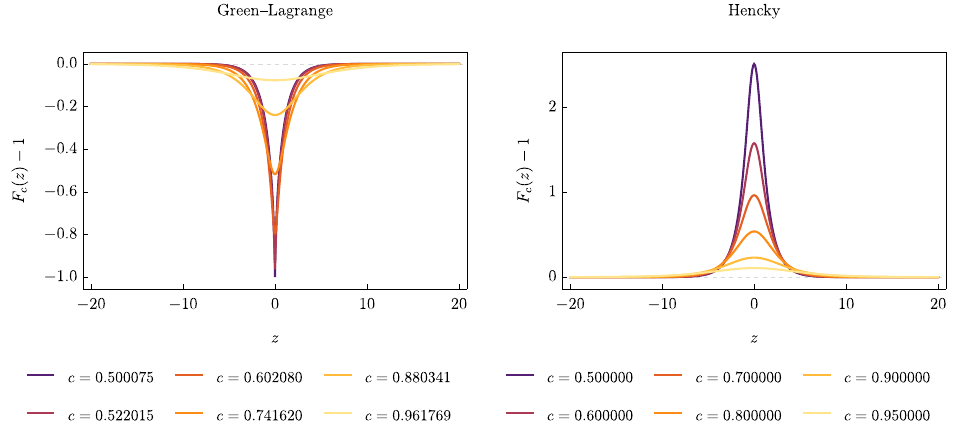}
	\caption{Admissible solitary wave profiles $F_c(z)-1$ for Green--Lagrange and Hencky strains with quadratic lower-order energy $\mathscr{U}(\eta(F))$ in dimensionless variables. The normalization is defined in Section~\ref{sec:numerical-illustrations}. For the Green--Lagrange strain, the plot was computed with an explicit parametric relation. For the Hencky strain, the plot was computed via quadrature. }
	\label{fig:quadratic_profiles}
\end{figure}

In both cases, the centered strain profile is determined for $z>0$ by
\begin{align}
	z=\sqrt{\frac{\kappa_\eta}{2}}
	\int_{|\varepsilon(z)|}^{A}
	\frac{da}{\sqrt{Q_c(\sigma a)}},
\end{align}
and extended evenly across the center. Proposition~\ref{general} supplies smoothness, positive stretch, exponential decay with rate $\sqrt{(\mu-\rho_0c^2)/\kappa_\eta}$, finite mechanical energy, and uniqueness up to translation within the corresponding strain segment. Quadratic spring energies therefore already support solitary waves through their dependence on a finite-strain measure and its gradient.

These applications establish existence and determine the profiles by parametric relations and quadrature. To obtain explicit profiles, we next adjust the spring energy so that a change of strain variable makes the quadrature explicitly invertible.
\FloatBarrier
\subsection{A strain transformation for explicit profiles}

We now seek a variable $r$ that makes the wave-speed term in \eqref{finite-strain-universal-first-integral} quadratic.
Define $r=R(\varepsilon)$ by
\begin{align}
	R(\varepsilon)
	=\varepsilon\exp\left[
	\int_0^\varepsilon
	\left(
	\frac{1}{\eta^{-1}(s)-1}-\frac{1}{s}
	\right)\,ds
	\right].
	\label{canonical-strain-map-integral}
\end{align}
The integrand in \eqref{canonical-strain-map-integral} has a removable
singularity at the origin. Moreover, $R(\varepsilon)$ has the same sign as
$\varepsilon$, which, because $\eta$ is strictly increasing and $\eta(1)=0$, has
the same sign as $F-1$. Thus, positive and negative values of $r$ represent
tensile and compressive states, respectively. The map also satisfies
\begin{align}
	R'(\varepsilon)(\eta^{-1}(\varepsilon)-1)=R(\varepsilon),
	\quad
	R(0)=0,
	\quad
	R'(0)=1.
	\label{canonical-strain-map-equation}
\end{align}

The strain variable $r$ simplifies the first integral \eqref{finite-strain-universal-first-integral} as follows. Let $\mathscr V$ be a smooth function satisfying
$\mathscr V(0)=\mathscr V'(0)=0$, and suppose that the lower-order energy
has the form
\begin{align}
	\mathscr U(\varepsilon)
	=\frac{\mathscr V(R(\varepsilon))}
	{(R'(\varepsilon))^2}.
	\label{weighted-pullback-energy}
\end{align}
Multiplication of
\eqref{finite-strain-universal-first-integral} by
$(R'(\varepsilon))^2$ then gives
\begin{align}
	\frac{\ka_\eta}{2}(r')^2
	=\mathscr V(r)-\frac{\rho_0c^2}{2}r^2.
	\label{transformed-polynomial-first-integral}
\end{align}
Indeed, $r'=R'(\varepsilon)\varepsilon'$ and
$R'(\varepsilon)(\eta^{-1}(\varepsilon)-1)=R(\varepsilon)=r$.
Moreover, since $R(0)=0$, $R'(0)=1$, and
$\mathscr V(0)=\mathscr V'(0)=0$, direct differentiation gives
\begin{align}
	\mathscr U''(0)=\mathscr V''(0).
\end{align}
Consequently,
\begin{align}
	\mathscr U''(0)=\mu
	\iff 
	\mathscr V''(0)=\mu.
\end{align}
Thus, the change of variable preserves the prescribed small-strain
modulus. Unlike the original first integral, the transformed equation \eqref{transformed-polynomial-first-integral} isolates
the dependence on the wave speed in a quadratic term and removes the
generally nonlinear expression involving $\eta^{-1}$.

The transformation also connects the explicit construction to Proposition~\ref{general}. Define
\begin{align}
	\widetilde Q_c(r):=\mathscr V(r)-\frac{\rho_0c^2}{2}r^2.
\end{align}
Then
\begin{align}
	Q_c(\varepsilon)=\frac{\widetilde Q_c(R(\varepsilon))}{(R'(\varepsilon))^2}.
\end{align}
The definition of $R$ implies that $R'>0$ on $K$. Thus $R$ is a smooth increasing map onto $R(K)$, and positivity and simple turning points transfer between $Q_c$ and $\widetilde Q_c$. In particular, a homoclinic orbit satisfying the transformed turning-point conditions gives the solitary wave of Proposition~\ref{general} whenever $\mu_c>0$ and the closed segment from zero to its turning point lies in $R(K)$. This range condition will determine the additional admissibility restrictions in the examples below.

\subsection{A quartic potential and its homoclinic profile}
As a representative example, we take
\begin{align}
	\mathscr V(r)
	=\frac{\mu}{2}r^2-\frac{a}{3}r^3+\frac{b}{4}r^4,
	\quad
	\mu>0,
	\quad a\neq0,
	\quad b>0.
	\label{transformed-quartic-potential}
\end{align}
Set
\begin{align}
	\sigma=\operatorname{sgn}(a).
\end{align}
For a given speed $c$, define
\begin{align}
	\mu_c :=\mu-\rho_0c^2,
	\quad
	\delta_c:=\sqrt{\frac{4a^2}{9}-2b\mu_c}.
	\label{transformed-profile-parameters}
\end{align}
Suppose that
\begin{align}
	\mu_c>0,
	\quad
	\frac{4a^2}{9}-2b\mu_c>0.
	\label{transformed-homoclinic-conditions}
\end{align}
Then, up to translation, the nontrivial homoclinic solution of
\eqref{transformed-polynomial-first-integral} is
\begin{align}
	r_c(z)
	=\sigma\frac{2\mu_c}
	{\dfrac{2|a|}{3}+\delta_c
	\cosh\left(\sqrt{\mu_c/\ka_\eta}\,(z-z_0)\right)},
	\quad z_0\in\bbR.
	\label{common-transformed-solitary-wave}
\end{align}
Indeed, a direct substitution shows that
\begin{align}
	\ka_\eta r_c''
	=\mu_c r_c-ar_c^2+br_c^3,
\end{align}
and that \eqref{transformed-polynomial-first-integral} holds. The function $r_c$ is positive when $a>0$ and negative when $a<0$. If
\begin{align}
	\lambda_{\eta,c}=\sqrt{\frac{\mu_c}{\ka_\eta}},
\end{align}
then
\begin{align}
	r_c(z)
	=\sigma\frac{4\mu_c}{\delta_c}
	\exp\bigl(-\lambda_{\eta,c}|z-z_0|\bigr)
	+O\left(
	\exp\bigl(-2\lambda_{\eta,c}|z-z_0|\bigr)
	\right)
	\label{transformed-profile-decay}
\end{align}
as $|z|\to\infty$. The maximum
absolute value of the profile is
\begin{align}
	\widehat r_*(c)
	:=|r_c(z_0)|
	=\frac{2\mu_c}{\dfrac{2|a|}{3}+\delta_c}
	=\frac{\dfrac{2|a|}{3}-\delta_c}{b}.
	\label{transformed-profile-amplitude}
\end{align}

Existence of the transformed orbit and positivity of the constitutive energy impose separate conditions. Requiring $\mathscr V(r)>0$ for every $r\ne0$ is equivalent to
\begin{align}
	2a^2<9b\mu
	\label{positive-transformed-potential-condition}
\end{align}
{because}
\begin{align}
\min_{r\in\bbR}
	\left(\frac{\mu}{2}-\frac{a}{3}r+\frac{b}{4}r^2\right)
	=\frac{\mu}{2}-\frac{a^2}{9b}.
\end{align}

The second condition in
\eqref{transformed-homoclinic-conditions} is equivalent to
\begin{align}
	2a^2>9b(\mu-\rho_0c^2).
\end{align}
Thus, positivity of $\mathscr V$ and existence of the
homoclinic profile are compatible when
\begin{align}
	9b(\mu-\rho_0c^2)<2a^2<9b\mu.
	\label{positive-energy-wave-regime}
\end{align}
It follows that the admissible speeds satisfy
\begin{align}
	\sqrt{\frac{\mu-2a^2/(9b)}{\rho_0}}
	<|c|<
	\sqrt{\frac{\mu}{\rho_0}}.
	\label{finite-strain-wave-speed-window}
\end{align}
In particular, these waves are subsonic relative to the infinitesimal longitudinal wave speed $\sqrt{\mu/\rho_0}$. This window ensures existence of the transformed orbit and positivity of the energy. To obtain an admissible physical wave, its amplitude must also remain in $R(K)$; we check this separately for each strain measure.

\begin{figure}[t]
	\centering
	\includegraphics[width=\linewidth]{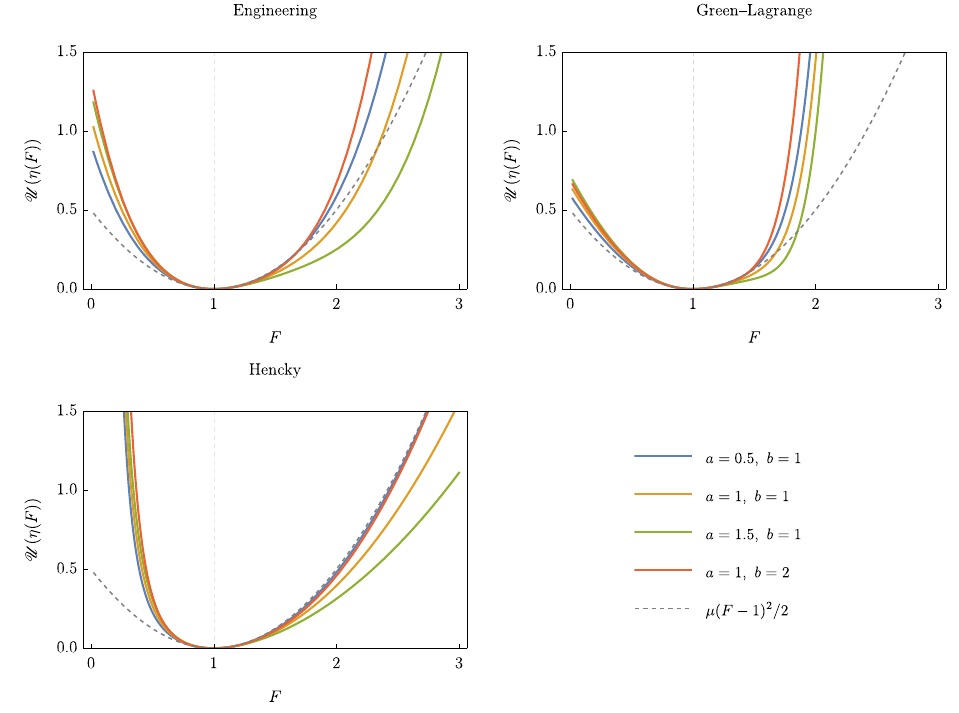}
	\caption{Dimensionless lower-order energy $\mathscr U(\eta(F))$ for the engineering,
	Green--Lagrange, and Hencky strains with $a>0$ and $\mu=1$. The dashed
	curve is the common quadratic $\mu(F-1)^2/2$. Every pair $(a,b)$ satisfies
	\eqref{positive-transformed-potential-condition}. The normalization is defined in Section~\ref{sec:numerical-illustrations}.}
	\label{fig:energy-tensile}
\end{figure}

\begin{figure}[t]
	\centering
	\includegraphics[width=\linewidth]{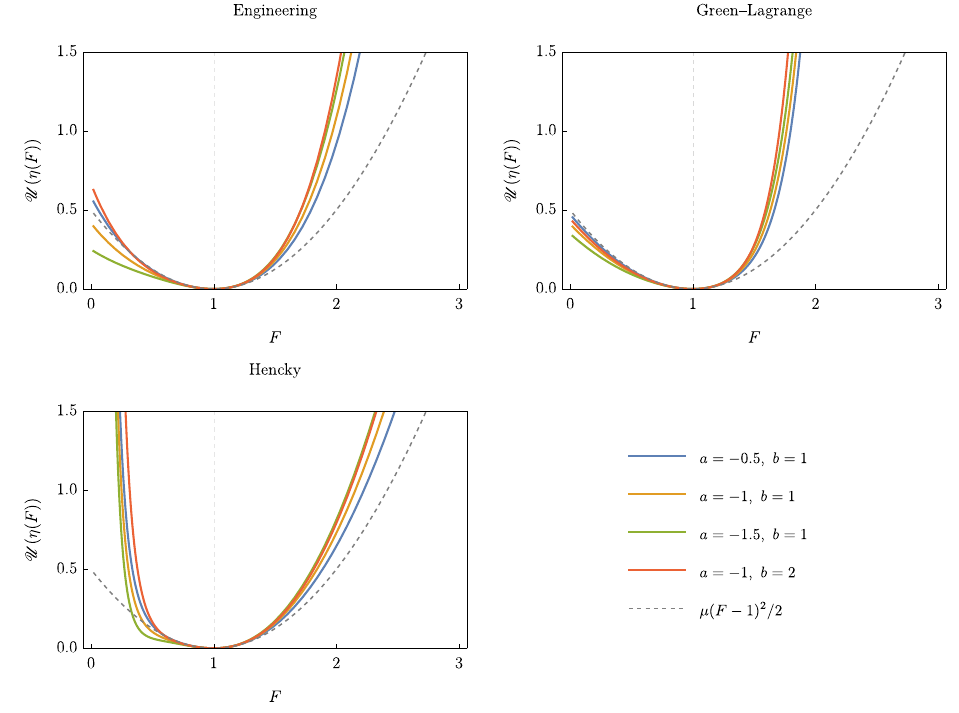}
	\caption{As in Figure~\ref{fig:energy-tensile}, with $a<0$. The same dimensionless normalization is used.}
	\label{fig:energy-compressive}
\end{figure}

\subsection{Engineering-strain model}

For the engineering strain $\eta(F)=F-1$, the matched energy
\eqref{solvable-matched-energy} becomes
\begin{align}
	\Psi_{\rm eng}(F,H)
	=\frac{\ka_H}{2}H^2+\mathscr U(F-1).
\end{align}
In this case, the canonical map is the identity:
\begin{align}
	r=R(\varepsilon)=\varepsilon=F-1.
\end{align}
Taking $J=(0,\infty)$ gives $K=R(K)=(-1,\infty)$. For the quartic potential \eqref{transformed-quartic-potential}, the spring energy is $\mathscr U(\varepsilon)=\mathscr V(\varepsilon)$; under \eqref{positive-transformed-potential-condition}, it is positive away from the natural state and has linearized modulus $\mu$. The physical profile is therefore
\begin{align}
	F_{{\rm eng},c}(z)=1+r_c(z),
\end{align}
with $\kappa_\eta=\kappa_H$ in \eqref{common-transformed-solitary-wave}. Tensile profiles are automatically admissible. Compressive profiles require $\widehat r_*(c)<1$ so that the entire strain segment remains in $(-1,\infty)$. Under \eqref{transformed-homoclinic-conditions} and this range restriction, Proposition~\ref{general} gives a unique solitary wave in the corresponding segment, with positive stretch and finite mechanical energy.

The engineering-strain model also connects the construction with a familiar evolution equation. If $e=F-1$, then the full dynamic
equation for the stretch, in the absence of body forces, is
\begin{align}
	\rho_0 e_{tt}
	=\partial_X^2\left(\mathscr U'(e)-\ka_H e_{XX}\right),
\end{align}
or, equivalently,
\begin{align}
	\rho_0 e_{tt}
	-\partial_X^2\mathscr U'(e)
	+\ka_H e_{XXXX}=0.
\end{align}
Thus, the complete stretch equation has
the form of a generalized good Boussinesq equation, independently of any
traveling wave ansatz.

The Hencky and Green--Lagrange examples considered next are different.
Their gradient moduli are
\begin{align}
	A_L(F)=\frac{\ka_L}{F^2},
	\quad
	A_G(F)=\ka_G F^2,
\end{align}
and their full dynamic equations contain the corresponding finite-strain
nonlinearities. We note that the change of variable to $r$ transforms the profile
first integral into
\eqref{transformed-polynomial-first-integral}, but it does not transform the
full evolution equation into a good Boussinesq equation. Instead, the
familiar homoclinic profile serves as an auxiliary orbit whose pullback
produces explicit solitary waves for two finite-strain second-gradient
energies with deformation-dependent gradient moduli.

\subsection{Green--Lagrange-strain model}

We now consider 
\begin{align}
	\eta(F)=E=\frac{F^2-1}{2},
	\quad
	\Psi_G(F,H)
	=\frac{\ka_G}{2}G^2+\mathscr U_G(E)
	=\frac{\ka_G}{2}F^2H^2+\mathscr U_G(E).
	\label{solvable-green-energy-general}
\end{align}
Introduce the stretch increment
\begin{align}
	s=F-1=\sqrt{1+2E}-1 \in (-1,\infty),
	\quad
	E=s+\frac{s^2}{2}.
	\label{green-stretch-increment}
\end{align}
Since $\eta^{-1}(E)-1=s$, equation
\eqref{canonical-strain-map-equation} becomes $R_G'(E)s=R_G(E)$. Using
$dE=(1+s)\,ds$, we obtain
\begin{align}
	r=R_G(E)=se^s,
	\quad
	R_G'(E)=e^s.
	\label{green-canonical-strain-map}
\end{align}
The corresponding lower-order energy is
\begin{align}
	\mathscr U_G(E)
	=e^{-2s(E)}\mathscr V\bigl(s(E)e^{s(E)}\bigr),
	\quad
	s(E)=\sqrt{1+2E}-1.
	\label{solvable-green-lower-order-energy-general}
\end{align}
For \eqref{transformed-quartic-potential}, this becomes
\begin{align}
	\mathscr U_G(E)
	=\frac{\mu}{2}s^2
	-\frac{a}{3}s^3e^s
	+\frac{b}{4}s^4e^{2s},
	\quad
	s=\sqrt{1+2E}-1.
	\label{explicit-green-lower-order-energy}
\end{align}
The expansion at the reference state is
\begin{align}
	\mathscr U_G(E)
	=\frac{\mu}{2}E^2
	-\left(\frac{\mu}{2}+\frac{a}{3}\right)E^3
	+\left(
	\frac{5\mu}{8}+\frac{a}{6}+\frac{b}{4}
	\right)E^4
	+O(E^5).
	\label{solvable-green-energy-expansion}
\end{align}
Under
\eqref{positive-transformed-potential-condition},
$\mathscr U_G(E)>0$ for every nonzero admissible $E$. The energy grows
without bound as $F\to\infty$, but it remains finite as $F\to0^+$.

We now pull back the common homoclinic orbit to obtain the physical wave. Under \eqref{transformed-homoclinic-conditions}, define
\begin{align}
	r_{G,c}(z)
	=\sigma\frac{2\mu_c}
	{\dfrac{2|a|}{3}+\delta_c
		\cosh\left(\sqrt{\mu_c/\ka_G}\,(z-z_0)\right)}.
	\label{explicit-green-transformed-profile}
\end{align}
The principal real branch of the Lambert function, denoted by
$\operatorname{W}_0$, is determined by
\begin{align}
	\operatorname{W}_0(r)e^{\operatorname{W}_0(r)}=r,
	\quad r\geq-e^{-1},\quad \operatorname{W}_0(r)\geq-1.
\end{align}
When $a<0$, assume in addition that
\begin{align}
	\widehat r_*(c)<e^{-1}.
	\label{green-profile-amplitude-condition}
\end{align}
No additional amplitude condition is needed when $a>0$. Under these
conditions, inversion of \eqref{green-canonical-strain-map} gives
\begin{align}
	s_c(z)
	&=\operatorname{W}_0\bigl(r_{G,c}(z)\bigr),
	\label{explicit-green-stretch-increment-profile}\\
	F_{G,c}(z)
	&=1+ s_c(z) = 1+\operatorname{W}_0\bigl(r_{G,c}(z)\bigr),
	\label{explicit-green-stretch-profile}\\
	E_c(z)
	&=\operatorname{W}_0\bigl(r_{G,c}(z)\bigr)
	+\frac12\operatorname{W}_0\bigl(r_{G,c}(z)\bigr)^2.
	\label{explicit-green-strain-profile}
\end{align}
The range of $R_G$ for $F>0$ is $(-e^{-1},\infty)$. Thus, for $a>0$ the
wave is tensile and $F_{G,c}>1$. For $a<0$, condition
\eqref{green-profile-amplitude-condition} keeps the profile above the
branch point $-e^{-1}$; the wave is compressive and satisfies
$0<F_{G,c}<1$.
The compressive amplitude condition is automatic when
$2|a|/3\leq b/e$. If $2|a|/3>b/e$, it is equivalent to
\begin{align}
	\mu_c<\frac{2|a|}{3e}-\frac{b}{2e^2}
	\iff 
	\rho_0c^2>
	\mu-\frac{2|a|}{3e}+\frac{b}{2e^2}.
	\label{green-profile-speed-condition}
\end{align}
These range restrictions place the turning point in $R_G(K)$ and therefore verify the hypotheses of Proposition~\ref{general} for the Green--Lagrange wave.

\subsection{Hencky-strain model}
\label{sec:hencky}
We now consider the case 
\begin{align}
	\eta(F)=L=\log F,
	\quad
	\Psi_L(F,H)
	=\frac{\ka_L}{2F^2}H^2+\mathscr U_L(L).
	\label{solvable-hencky-energy-general}
\end{align}
In this case
\begin{align}
	\eta^{-1}(L)-1=e^L-1.
\end{align}
The normalized solution of \eqref{canonical-strain-map-equation} is
\begin{align}
	r=R_L(L)=1-e^{-L}=1-\frac{1}{F},
	\quad
	R_L'(L)=e^{-L}.
	\label{hencky-canonical-strain-map}
\end{align}
It follows from \eqref{weighted-pullback-energy} that
\begin{align}
	\mathscr U_L(L)
	=e^{2L}\mathscr V(1-e^{-L}).
	\label{solvable-hencky-lower-order-energy}
\end{align}
For the quartic potential
\eqref{transformed-quartic-potential}, the lower-order contribution may be
written directly in terms of $F$ as
\begin{align}
	\mathscr U_L(\log F)
	=\frac{\mu}{2}(F-1)^2
	-\frac{a}{3}\frac{(F-1)^3}{F}
	+\frac{b}{4}\frac{(F-1)^4}{F^2}.
	\label{explicit-hencky-lower-order-energy}
\end{align}
Near the reference state, the energy has the expansion
\begin{align}
	\mathscr U_L(L)
	=\frac{\mu}{2}L^2
	+\left(\frac{\mu}{2}-\frac{a}{3}\right)L^3
	+\left(
	\frac{7\mu}{24}-\frac{a}{6}+\frac{b}{4}
	\right)L^4
	+O(L^5).
	\label{solvable-hencky-energy-expansion}
\end{align}
If
\eqref{positive-transformed-potential-condition} holds, then
$\mathscr U_L(L)>0$ for $L\neq0$, and moreover,
\begin{align}
	\mathscr U_L(\log F)
	&= \frac{b}{4F^2} + O(F^{-1})
	\quad \text{as }F\to0^+,\notag\\
	\mathscr U_L(\log F)
	&=\mathscr V(1)F^2+O(F)
	\quad \text{as }F\to\infty.
	\label{solvable-hencky-energy-coercivity}
\end{align}
Under condition
\eqref{positive-transformed-potential-condition}, the lower-order
energy diverges both as the stretch approaches zero and as it becomes
unbounded. Thus, the material strongly resists both collapse and unlimited extension. Together with
the identities
\begin{align}
	\mathscr U_L(0)=\mathscr U_L'(0)=0,
	\quad
	\mathscr U_L''(0)=\mu>0,
\end{align}
and the strict positivity of $\mathscr U_L(L)$ for $L\neq0$, these
asymptotics show that {$F=1$ is the unique zero and global
minimizer of the lower-order energy and that the reference state has a
positive linearized modulus.} In
addition, the gradient contribution
\begin{align}
	\frac{\ka_L}{2F^2}H^2
\end{align}
is nonnegative and vanishes precisely for locally homogeneous
deformations. Consequently, the complete stored-energy density
$\Psi_L(F,H)$ is nonnegative, vanishes only at the natural state
$(F,H)=(1,0)$, penalizes spatial variations of stretch, and becomes
unbounded under collapse, unlimited extension, or unbounded strain
gradients. In the sense of the properties listed above, the
constructed energy $\Psi_L$ is a \textit{physically natural}
finite-strain second-gradient energy for future mathematical study. In fact, the following proposition shows that finite mechanical energy provides quantitative control of the stretch throughout any energy-conserving smooth evolution (for the proof see Appendix \ref{sec:AppendixB}).
\begin{prop}
\label{prop:stored_energy_coercivity} Suppose $\mu, b, \kappa_L >0, 2a^2 < 9b\mu$, and $\ell>0$ is a fixed length scale. For the Hencky stored-energy density 
   \[\Psi_L(F, H) = \frac{\kappa_L}{2F^2}H^2 + \mathscr U_L(L),\] there exists a $C>0$ such that every continuously differentiable positive function $F$ with finite stored energy satisfies
   \begin{align}
   \label{eq:hencky_coercivity}
   \int_\bbR\Psi_L(F, F_X)\, dX  \geq C \|\log F\|_{H^1}^2,
   \end{align}
   where $\| f \|_{H^1}^2 := \int_\bbR (\ell^2 |f_X|^2 + |f|^2) dX$.
   Consequently, if $U$ is a classical unforced finite-energy solution to  \eqref{dynamic-equation} on $[0, T]$ then
   \begin{align}
   \label{eq:Hencky_compressive_extension_bound}
       \forall (X, t) \in \bbR \times [0,T], \quad e^{-M} \leq F(X, t) \leq e^M
   \end{align}
   where $M = \sqrt{\mathcal{E}_0/2\ell C}$ and $\mathcal{E}_0$ denotes the initial mechanical energy of $U$,
   \begin{align}
    \mathcal E_0 := \int_{\bbR} \bigl [ \frac{\rho_0}{2} |\dot U(X, 0)|^2 + \Psi_L(F(X,0), F_X(X,0))\bigr ] dX. 
   \end{align}
\end{prop}

We now turn to the explicit solitary-wave solutions associated to $\mathscr U_L(L)$. Assume \eqref{transformed-homoclinic-conditions} and set
\begin{align}
	r_{L,c}(z)
	=\sigma\frac{2\mu_c}
	{\dfrac{2|a|}{3}+\delta_c
	\cosh\left(\sqrt{\mu_c/\ka_L}\,(z-z_0)\right)}.
	\label{explicit-hencky-transformed-profile}
\end{align}
If $a>0$, assume in addition that
\begin{align}
	\widehat r_*(c)<1.
	\label{hencky-profile-amplitude-condition}
\end{align}
No additional amplitude condition is needed when $a<0$. In either case,
inversion of \eqref{hencky-canonical-strain-map} gives the solitary wave
\begin{align}
	F_{L,c}(z)
	&=\frac{1}{1-r_{L,c}(z)},
	\label{explicit-hencky-stretch-profile}\\
	L_c(z)
	&=-\log\bigl(1-r_{L,c}(z)\bigr).
	\label{explicit-hencky-strain-profile}
\end{align}
Indeed, the range of $R_L$ is $(-\infty,1)$. Thus, condition
\eqref{hencky-profile-amplitude-condition} ensures that the tensile profile
for $a>0$ remains in this range. When $a<0$, one has $r_{L,c}<0$ and hence
\begin{align}
	0<F_{L,c}(z)<1,
\end{align}
so the resulting wave is compressive and automatically admissible.
The tensile amplitude condition is automatic when $2|a|/3\leq b$. If
$2|a|/3>b$, it is equivalent to
\begin{align}
	\mu_c<\frac{2|a|}{3}-\frac{b}{2},
	\iff
	\rho_0c^2>\mu-\frac{2|a|}{3}+\frac{b}{2}.
	\label{hencky-profile-speed-condition}
\end{align}
These restrictions likewise place the turning point in $R_L(K)$, and Proposition~\ref{general} establishes the qualitative properties of the Hencky wave.

\begin{figure}[t]
	\centering
	\includegraphics[width=\linewidth]{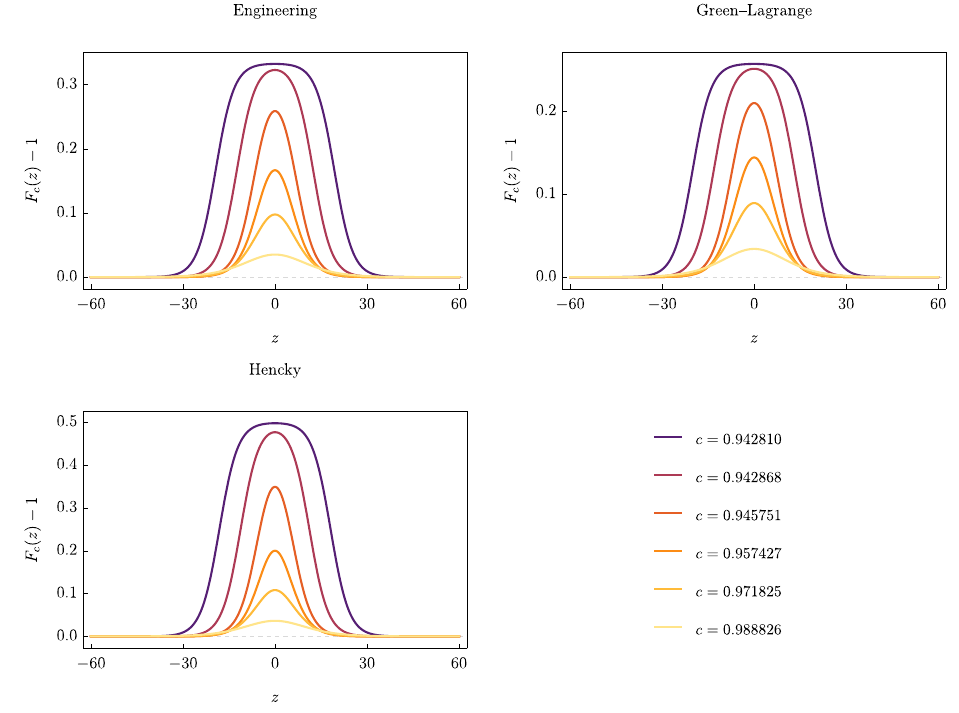}
	\caption{Tensile solitary-wave profiles $F_c(z)-1$ in dimensionless variables for $a=1$, $b=2$,
	$\rho_0=\mu=\ka_\eta=1$, and speeds within the admissible window $\sqrt{8}/3 < c < 1$.}
	\label{fig:profile-tensile}
\end{figure}

\begin{figure}[t]
	\centering
	\includegraphics[width=\textwidth]{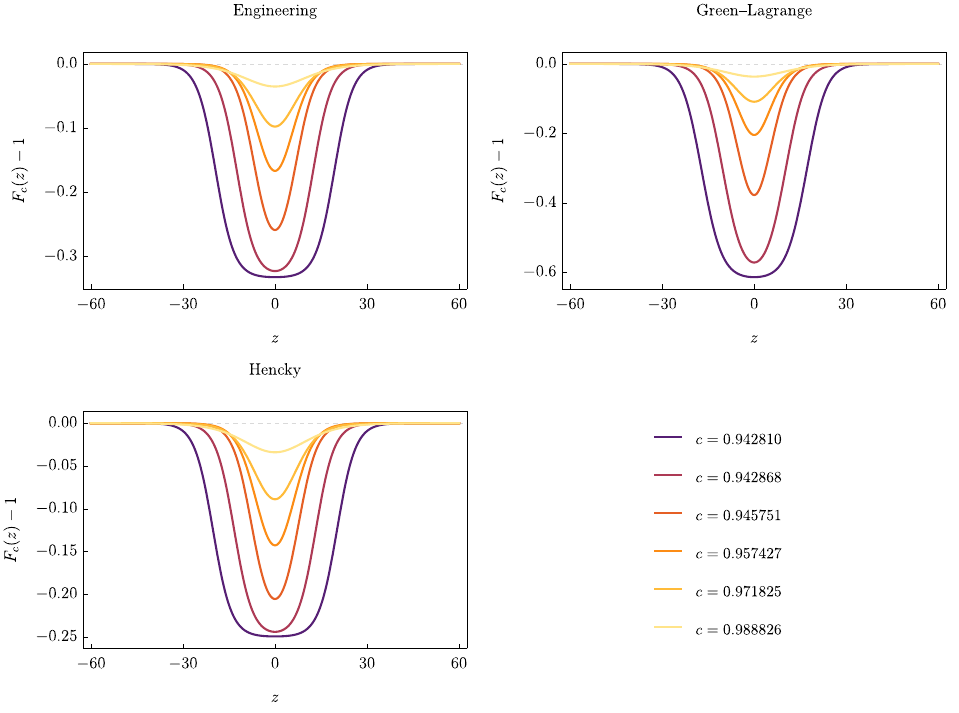}
	\caption{Compressive solitary-wave profiles $F_c(z)-1$ in dimensionless variables for $a=-1$, $b=2$,
	$\rho_0=\mu=\ka_\eta=1$, and speeds within the admissible window $\sqrt{8}/3 < c< 1$.}
	\label{fig:profile-compressive}
\end{figure}

\subsection{Decay of the explicit profiles}

For all three strain measures, $R(0)=0$ and $R'(0)=1$. Combining these normalizations with \eqref{transformed-profile-decay} gives the precise physical tails
\begin{align}
	F_c(z)-1
	=\sigma\frac{4\mu_c}{\delta_c}
	\exp\bigl(-\lambda_c|z-z_0|\bigr)
	+O\left(\exp\bigl(-2\lambda_c|z-z_0|\bigr)\right),
	\label{physical-profile-decay}
\end{align}
where
\begin{align}
	\lambda_c=\sqrt{\mu_c/\kappa_\eta},
	\quad
	\kappa_\eta=
	\begin{cases}
		\kappa_H,&\text{for engineering strain},\\
		\kappa_G,&\text{for Green--Lagrange strain},\\
		\kappa_L,&\text{for Hencky strain}.
	\end{cases}
\end{align}
The explicit construction thus identifies the tail coefficient left unspecified in Proposition~\ref{general}. Once the model-specific amplitude restrictions hold, that proposition gives finite stored and kinetic energy and a finite nonzero displacement offset \eqref{finite-strain-displacement-shift}, positive for tensile waves and negative for compressive waves.

\subsection{Numerical illustrations}
\label{sec:numerical-illustrations}

For the figures, we introduce the dimensionless variables and parameters
\begin{align*}
\widehat z=z\sqrt{\frac{\mu}{\kappa_\eta}},\quad
\widehat c=c\sqrt{\frac{\rho_0}{\mu}},\quad
\widehat{\mathscr U}=\frac{\mathscr U}{\mu},\quad
\widehat a=\frac{a}{\mu},\quad
\widehat b=\frac{b}{\mu}.
\end{align*}
Here $\kappa_\eta$ is the gradient coefficient for the corresponding strain measure. In these variables the profile equations have normalized coefficients $\rho_0=\mu=\kappa_\eta=1$. Hats are omitted in the figures, their captions, and the discussion below; $F$ and $F-1$ are already dimensionless.

Figure~\ref{fig:quadratic_profiles} illustrates the solitary wave profiles $F_c(z) -1$ for the Green--Lagrange and Hencky strains with quadratic lower-order energies.  For the Green--Lagrange strain, the solitary wave was computed directly from the explicit parameterization \eqref{eq:GL_parameterization}. For the Hencky strain, the quadrature was evaluated with the substitution $L = A\sech^2(\xi)$, which removes the inverse-square-root singularity at the crest and converts the logarithmic divergence as $L \to 0$ into linear growth in $\xi$. The resulting smooth integral was computed numerically, and the solitary wave was recovered by inversion.

Figures~\ref{fig:energy-tensile}--\ref{fig:profile-compressive} illustrate
the three families for the quartic potential
\eqref{transformed-quartic-potential}, with
 $z_0=0$.
Figures~\ref{fig:energy-tensile} and~\ref{fig:energy-compressive} show the
lower-order energy \eqref{weighted-pullback-energy}. 
Figures~\ref{fig:profile-tensile} and~\ref{fig:profile-compressive} show
the solitary-wave profiles $F_c(z)-1$ for $a=\pm1$ and $b=2$, with $c$
ranging across the admissible window \eqref{finite-strain-wave-speed-window},
$\sqrt{8}/3<c<1$. For these parameters
$\widehat r_*(c)<2|a|/(3b)=1/3<e^{-1}$, so the amplitude conditions
\eqref{green-profile-amplitude-condition} and
\eqref{hencky-profile-amplitude-condition} hold for every admissible speed.
As $c$ approaches the lower limit, $\delta_c\to0$ and the profiles broaden
into flat-topped plateaus; as $c$ approaches the upper limit, they become
small and wide.
\FloatBarrier

\section{Conclusions and open questions}

One-dimensional second-gradient elasticity provides a concrete setting in which to connect generalized continuum mechanics with nonlinear dispersive PDEs. Starting from finite-strain kinematics and the Principle of Virtual Power, we have identified internal forces and contact actions, derived the governing equation, and established the mechanical-energy identity. For pantograph-inspired energies, the traveling-wave first integral gives a general existence criterion for admissible solitary waves. The quadratic-energy examples and explicit constructions show how finite-strain constitutive choices affect solitary-wave existence while leaving the homogeneous linear modulus unchanged; when the strain-gradient coefficients are also matched, the full linearized equation is unchanged. The resulting formulation and examples suggest several questions for further study.

The formulation and solitary waves developed here lead to two related mathematical questions. The first concerns well-posedness and small-data global existence: is the initial-value problem on the whole line locally well-posed for initial data with positive stretch and globally well-posed for small localized perturbations of the natural state? Addressing this question requires determining whether the mechanical-energy identity, together with suitable dispersive estimates, provides sufficient control of the nonlinear evolution. Proposition~\ref{prop:stored_energy_coercivity} provides a first step in this direction by establishing uniform upper and positive lower bounds on the stretch of smooth, energy-conserving solutions. Establishing global well-posedness requires additional estimates controlling higher spatial derivatives.

A well-posed evolution would also provide the setting for studying the stability of the tensile and compressive solitary waves. Which waves are spectrally or orbitally stable, and how does stability depend on speed, strain measure, and constitutive parameters? The explicit profiles permit direct study of the linearized operators, while the general existence criterion extends the scope of this question beyond the explicitly solvable families. Establishing nonlinear stability, or identifying mechanisms of instability, would clarify which localized motions persist under perturbations.

These analytical questions also motivate numerical investigation, for which the explicit waves provide reference solutions against which to assess approximations of the nonlinear dynamics. Recent comparisons of finite-element formulations for one- and two-dimensional strain-gradient elasticity in \cite{SararYildizdagFabbrocinoAbali2025} offer a useful computational baseline. Extending such comparisons to the present finite-strain dynamical setting would require assessing both spatial discretization accuracy and the fidelity of the discrete mechanical-energy balance. The explicit profiles would then provide benchmarks for numerical studies of wave propagation and stability.

Two mechanical extensions would change the structure of the governing PDEs and open further analytical questions. The first is to include kinetic energies depending on spatial gradients of velocity, as suggested by microstructural inertia \cite{Toupin1964, Mindlin1964, Germain1973}. The Principle of Virtual Power provides a framework for identifying both bulk inertial forces and inertial boundary actions and establishing the corresponding mechanical-energy identity. For example, a quadratic contribution in the velocity gradient produces a mixed space-time derivative in the equation of motion. Its interaction with the elastic gradient term changes dispersion and the traveling-wave reduction, raising questions about admissible wave speeds and the existence of localized profiles. Deformation-dependent inertia would introduce an additional coupling between kinematics, constitutive response, and acceleration.

The second extension is to longitudinal motion of one-dimensional continua whose stored energies depend on third and higher spatial derivatives of placement. Their virtual-power formulations involve additional generalized forces and boundary actions, and their traveling-wave equations generally retain more degrees of freedom after integration. It remains to determine which constitutive structures permit useful first integrals, existence criteria, or explicit localized solutions. The resulting PDEs combine an elegant variational structure with a direct mechanical interpretation, linking questions of analysis to the response of materials with higher-order interactions.

\appendix

\section{Proof of Proposition \ref{general}}
\label{sec:appendixA}
Define $h(\varepsilon):=\eta^{-1}(\varepsilon)-1$ on $K$.
Since $h(0)=0$ and $h'(0)=1$, Taylor expansion gives
\begin{equation}
	Q_c(0)=Q_c'(0)=0,\quad Q_c''(0)=\mu_c,
	\quad q(s)=\frac{\mu_c}{2}s^2+O(s^3).
	\label{taylor}
\end{equation}
For $0<s<A$ define the quadrature
\begin{equation}
	Z(s)=\sqrt{\frac{\kappa_\eta}{2}}\int_s^A\frac{dt}{\sqrt{q(t)}}.
	\label{quad}
\end{equation}
The simple zero at $A$ gives
$q(t)=-q'(A)(A-t)+O((A-t)^2)$. Thus the integral is finite at its
upper endpoint and $Z(s)\to0$ as $s\rar A^-$. At the other endpoint,
\eqref{taylor} implies
\[
\sqrt{\frac{\kappa_\eta}{2q(s)}}=\frac{1}{\lambda_{\eta,c} s}+O(1).
\]
Consequently $Z(s)\to\infty$ as $s\rar 0^+$. Moreover,
$Z'(s)=-\sqrt{\kappa_\eta/(2q(s))}<0$. It follows that $Z$ is a smooth
bijection from $(0,A)$ onto $(0,\infty)$. Let $s(z)$ be its inverse for
$z>0$ and set $s(0)=A$. Differentiating the inverse yields
\begin{equation}
	\forall z \in (0, \infty), \quad s'=-\sqrt{\frac{2q(s)}{\kappa_\eta}},\quad
	\kappa_\eta s''=q'(s).
	\label{sode}
\end{equation}
In particular, $s(z) \rar 0^+$ as $z\to\infty$ and
$s'(z)\to0$ as $z\rar0^+$.

We justify smooth reflection at the crest, rather than relying on a
formal choice of sign in the first-order equation. Set $D=-q'(A)>0$.
The expansion of \eqref{quad} near $A$ gives
\[
Z(s)=\sqrt{\frac{2\kappa_\eta}{D}}\sqrt{A-s}\,(1+O(A-s)).
\]
Hence $A-s(z)=Dz^2/(2\kappa_\eta)+o(z^2)$, and
\eqref{sode} shows that the right derivatives at zero satisfy
$s'(0)=0$, $s''(0)=q'(A)/\kappa_\eta$. Thus $s$ extends to a $C^2$
solution of $\kappa_\eta s''=q'(s)$ on a right neighborhood of zero.
The smooth second-order equation with initial data $(s,s')=(A,0)$
has a unique smooth local solution. By uniqueness that solution equals
the quadrature solution on the right, and time reversal makes it even.
The reflection $s(-z)=s(z)$ is therefore smooth at zero and solves
the same equation everywhere. This also proves that
$s''(0)<0$, so the crest is a genuine turning point.

Set $\varepsilon=\sigma s$. Since
$q'(s)=\sigma Q_c'(\sigma s)$, we have on all of $\mathbb R$
\begin{equation}
	\kappa_\eta\varepsilon''=Q_c'(\varepsilon)
	=\mathscr{U}'(\varepsilon)-\rho_0c^2 h(\varepsilon)h'(\varepsilon).
	\label{epsode}
\end{equation}
The first integral holds for $z\ne0$ by construction and at zero by
continuity. In particular, it holds across the crest, with no singularity.
The range of $\varepsilon$ lies in a compact subset of $K$; consequently
$f=\eta^{-1}(\varepsilon)$ is smooth and takes values in a compact
subset of $J\subset(0,\infty)$.

To identify the tails precisely, subtract the singular integrand in
\eqref{quad}. Its remaining part is bounded at zero, and integrable at
$A$. Therefore, for some real constant $B$,
\[
Z(s)=-\frac{1}{\lambda_{\eta,c}}\log s+B+O(s)
\quad(s\rar0^+).
\]
Inverting gives $s(z)=C e^{-\lambda_{\eta,c} z}+O(e^{-2\lambda_{\eta,c} z})$ on the
right, where $C=e^{\lambda_{\eta,c} B}>0$. Reflection gives \eqref{tails}.
Also \eqref{sode} and \eqref{taylor} give
$s'=O(e^{-\lambda_{\eta,c} z})$, $s''=O(e^{-\lambda_{\eta,c} z})$.
Since $h(\varepsilon)=\varepsilon+O(\varepsilon^2)$ and $h$ has
bounded derivatives on the profile range, the asserted limits and
decay for $f-1,f',f''$ follow.

We now verify the original equation. The inverse-function identity is
$h'(\varepsilon)=1/\eta'(f)$. Multiplying \eqref{epsode} by
$\eta'(f)$ gives
\begin{equation}
	\eta'(f)\bigl(\mathscr{U}'(\varepsilon)-\kappa_\eta\varepsilon''\bigr)
	=\rho_0c^2(f-1).\label{tension}
\end{equation}
Direct differentiation of \eqref{solvable-matched-energy} yields along the profile
\[
\Psi_F-\partial_z\Psi_H
=\eta'(f)\bigl(\mathscr{U}'(\varepsilon)-\kappa_\eta\varepsilon''\bigr).
\]
Define
\[
\mathcal U(z)=u_-+\int_{-\infty}^z(f(y)-1)\,dy,
\quad U(X,t)=\mathcal U(X-ct).
\]
The integral converges by the exponential tails. Then
$U_X=f-1$, $\ddot U=c^2f'$, and differentiating \eqref{tension}
proves \eqref{dynamic-equation}. The motion $\chi=X+U$ is orientation preserving
and injective because $\chi_X=f>0$. In fact it maps $\mathbb R$ onto $\mathbb R$,
since $U$ is bounded. The two displacement limits differ by
$\int_\mathbb R(f-1)\,dz$, which has sign $\sigma$ and is nonzero because
$f-1$ has that strict sign at every finite point.

For the energy, $\mathscr{U}(\varepsilon)=O(\varepsilon^2)$ near zero, and
both $(f-1)^2$ and $(\varepsilon')^2$ decay like
$O(e^{-2\lambda_{\eta,c}|z|})$. Thus every term in \eqref{total} is absolutely
integrable on the tails and bounded on compact intervals. Along the
wave, $\eta'(f)f'=\varepsilon'$, so this is exactly the mechanical
energy of \eqref{solvable-matched-energy}, including the kinetic contribution
$U_t=-c(f-1)$.

Finally, consider any nonconstant classical solitary wave with strain
in the stated segment. Integrating its equation of motion and using
the solitary-wave limits gives \eqref{tension}; multiplying the
resulting \eqref{epsode} by $\varepsilon'$ and using the limits again
gives \eqref{finite-strain-universal-first-integral}. Such a wave cannot reach $\varepsilon=0$ at a
finite point: \eqref{finite-strain-universal-first-integral} would give $\varepsilon'=0$ there, and
uniqueness for \eqref{epsode} would force the zero solution. The
continuous function $|\varepsilon|$, tending to zero at both ends,
attains a positive maximum at a finite point. Its derivative vanishes
there, so \eqref{turning} and \eqref{finite-strain-universal-first-integral} force that maximum to be
$A$. At that point the initial data for \eqref{epsode} are
$(\varepsilon,\varepsilon')=(\sigma A,0)$. Uniqueness of that equation
now gives the constructed profile up to translation. Integrating
$\mathcal U'=f-1$ leaves only an additive displacement constant.
\qed 

\section{Proof of Proposition~\ref{prop:stored_energy_coercivity}}
\label{sec:AppendixB}
Define $\mathcal{W}[F] := \int_\bbR \Psi_L(F, F_X) dX$. For $X \in \bbR, t \in [0, T]$, let $L(X,t) = \log F(X,t)$. Under our parameter constraints and the estimates proved in Section~\ref{sec:hencky}, it follows that there exists a $c >0$ such that $\mathscr{U}_L(L) \geq \frac{1}{2} c L^2$ for all $L \in \bbR$, and thus
\begin{align}
\Psi_L(F, F_X) = \frac{\kappa_L}{2} |L_X|^2 + \mathscr U_L(L) \geq \frac{1}{2}\min(\kappa_L \ell^{-2}, c ) 
\left [\ell^2 |L_X|^2 + |L|^2 \right ],
\end{align}
where $\ell$ is the fixed length scale appearing in the definition of $H^1$. This proves \eqref{eq:hencky_coercivity}. 
The Sobolev inequality $\|L\|_{L^\infty}^2 \leq \frac{1}{2\ell} \|L\|_{H^1}^2$ then implies 
\begin{gather}
    2\ell C\|\log F(\cdot, t)\|_{L^\infty}^2 \leq  \mathcal{W}[F(\cdot, t)] \leq \mathcal{E}(t),
\end{gather}
where $\mathcal{E}(t)$ denotes the total mechanical energy at time $t$. Since there are no external forces, energy is conserved for classical finite energy motions, so that $\mathcal{E}(t) = \mathcal{E}_0$ for all $t \in [0, T]$. Setting $M = \sqrt{\mathcal{E}_0/2\ell C}$ yields \eqref{eq:Hencky_compressive_extension_bound}.
\qed
\bibliographystyle{abbrv}
\bibliography{literature}

\bigskip

\footnotesize

\noindent \textsc{Department of Mathematics, The University of North Carolina at Chapel Hill, CB 3250 Phillips Hall,
	Chapel Hill, NC 27599}\\
\noindent \textit{E-mail address}: \texttt{scott\_hallyburton@unc.edu}

\bigskip 

\noindent \textsc{Dipartimento di Ingegneria Civile, Edile-Architettura e Ambientale, Universit\`a dell'Aquila, Italy}\\
\noindent \textit{E-mail address}: \texttt{ivan.giorgio@univaq.it}

\bigskip 

\noindent \textsc{Department of Mathematics, The University of North Carolina at Chapel Hill, CB 3250 Phillips Hall,
		Chapel Hill, NC 27599}\\
\noindent \textit{E-mail address}: \texttt{crodrig@email.unc.edu}

\end{document}